\documentclass{aa}  

\usepackage{graphicx}
 \usepackage{caption}
\usepackage{morefloats}
\usepackage{txfonts}
\usepackage{appendix}
\usepackage[usenames]{xcolor}
\newcommand{\J}{\mathrm{J}}
\newcommand{\Ks}{\mathrm{K_s}}
\begin{document}

   \title{Mapping the stellar age of the Milky Way bulge with the VVV.  II.
   \thanks{Based on observations made with the VISTA telescope at the La Silla Paranal Observatory, under the ESO programme ID 179.B-2002.}}

   \subtitle{  Deep JK$_s$ catalogs release based on PSF photometry}

   \author{F. Surot,
          \inst{1}
    E. Valenti,
         \inst{2,3}
    S. L. Hidalgo,
         \inst{4,5}
    M. Zoccali,
         \inst{1,6}
    O. A. Gonzalez,
          \inst{7}
    E. S{\"o}kmen,
         \inst{4,5}
   D. Minniti,
         \inst{6,8,9}    
   M. Rejkuba,
        \inst{2}
   P. W. Lucas
       \inst{10}
          }
\institute{Instituto de Astrof\'{i}sica, Pontificia Universidad Cat\'{o}lica de Chile, Av. Vicu\~{n}a Mackenna 4860, Santiago , Chile.\\
\email{frsurot@uc.cl}
\and          
European Southern Observatory, Karl Schwarzschild\--Stra\ss e 2, D\--85748 Garching \\
bei M\"{u}nchen, Germany. 
\email{evalenti@eso.org}
\and
Excellence Cluster ORIGINS, Boltzmann\--Stra\ss e 2, D\--85748 Garching bei M\"{u}nchen, Germany
\and
Instituto de Astrofisica de Canarias, Via Lacteas S/N, E\--38200, La Laguna Tenerife, Spain
\and
Department of Astrophisics, University of La Laguna, E\--38200, La Laguna Tenerife, Canary Islands, Spain
\and
Millennium Institute of Astrophysics, Av. Vicu\~{n}a Mackenna 4860, 782-0436 Macul, Santiago, Chile.
\and
UK Astronomy Technology Centre, Royal Observatory, Blackford Hill, Edinburgh, EH9 3HJ, UK
\and
Departamento de Ciencias Fis\'{i}cas, Universidad Andr\'{e}s Bello, Rep\'{u}blica 220, Santiago, Chile
\and
Vatican Observatory, V00120 Vatican City State, Italy
\and
Centre for Astrophysics Research, School of Physics, Astronomy and Mathematics, University of Hertfordshire, College Lane, Hatfield AL10 9AB, UK
}

   \date{}

 
  \abstract
   {
   The bulge represents the best compromise between old and massive Galactic component, and as such its study is a valuable opportunity to understand how the bulk of the Milky Way formed and evolved. 
   In addition, being the only bulge in which we can individually resolve stars in all evolutionary sequences, the properties of its stellar content provide crucial insights on the formation of bulges at large.
   }
   {
   We aim at providing a detailed and comprehensive census of the Milky Way bulge stellar populations by producing deep and accurate photometric catalogs of the inner $\sim300\,\mathrm{deg}^2$ of the Galaxy. 
   }
   {
   We perform DAOPHOT/ALLFRAME PSF\--fitting photometry of multi\--epochs J and K$_s$ images provided by the VISTA Variables in the V\'{i}a L\'{a}ctea (VVV) survey to obtain deep photometric catalogs. Artificial star experiments have been conducted on all images to properly assess the completeness and the accuracy of the photometric measurements.
      }
  {
We present a photometric database containing nearly 600 million stars across the bulge area surveyed by the VVV.  
Through the comparison of derived color\--magnitude diagrams of selected fields representative of different levels of extinction and crowding, we show the quality, completeness and depth of the new catalogs. With the exception of the fields located along the plane, this new photometry samples stars down to $\sim$\,1\--2\,mag below the old Main Sequence turnoff with unprecedented accuracy.
To demonstrate the tremendous potential inherent to this new dataset, we give few examples of possible applications such as: {\it i)} star counts studies through the dataset completeness map;  {\it ii)} surface brightness map; and {\it iii)} cross\--correlation with Gaia DR2.
  }
   {
The database presented here represents an invaluable collection for the whole community, and we encourage its exploitation. The photometric catalogs including completeness information are publicly available through the ESO Science Archive.
   }

    \authorrunning{Surot et al.}  

   \maketitle
%

\section{Introduction}
\label{sec:intro}

With a stellar mass of $2.0\pm0.3\times10^{10}\,\mathrm{M}_\odot$ \citep{valenti+16} the Milky Way bulge represents the most massive component hosting very old population ($>$\,10\,Gyr). As such, the detailed study of its stellar content in terms of 3D structure, metallicity, kinematics and age provides the best way to understand how the bulk of the Milky Way formed.
Because of its vast extension on the sky (i.e. $\sim$\,500 \,$\mathrm{deg}^2$) and the presence of large and highly differential extinction  (i.e. $0.3 \lesssim A_V \lesssim 40$),  near\--IR wide field observations are best suited to obtain a global view of the Milky Way bulge.

In this respect, the near\--IR VISTA Variables in the V\'{i}a L\'{a}ctea \citep[VVV,][]{minniti_vvv} survey represents a true gold mine, as it enables the study of most of the bulge stellar content, even in the innermost central regions, so far poorly explored.
The VVV is an ESO public survey started in 2010 and completed 5 years later. It provides multi\--band (i.e. Z,\,Y,\,J,\,H,\,K$_s$) and multi\--epoch (i.e. up to 293 times) photometric observations of about 300\,deg$^2$ centered on the Milky Way bulge, between $-10^\circ\leq l \leq+10.4^\circ$ and $-10.3^\circ\leq b \leq+5.1^\circ$. 

The first public data release (VVV\--DR1) based on aperture photometry was presented by \citet{saitovvv}, and it enabled a number of interesting results among which the first complete bulge extinction map by \citet{oscar12}, the 3D bulge structure model by \citet{wegg+13} and the proper motion and parallax catalog by \citet{VIRAC+18} deserve special mention. 
However, a serious limitation of the VVV\--DR1 is the photometric accuracy and completeness in the innermost regions because of the large stellar crowding. 
To overcome this problem, recently a new public release based on point spread function (PSF) fitting photometry of 2 VVV epochs in all five passbands, has been presented by \citet{alonso_vvv}. 
The first internal\footnote{initially only available within the VVV team} release of this dataset has been already extensively used in different studies. 
Some examples are: {\it i)} the red clump (RC) stellar density map and the first empirical estimate of the bulge stellar mass by \citet{valenti+16};  {\it ii)} the characterization of a spiral structure behind the Galactic bar using RC stars \citep{gonzalez+18}; and {\it iii)} the measurement of the extinction law towards the bulge innermost $\sim$\,9\,deg$^2$ by \citet{alonsogarcia+17} and in the direction of other VVV fields in common with OGLE by \citet{nataf+16}. 

The Dophot\--based PSF\--fitting photometry published by \citet{alonso_vvv} is deep enough to sample the Main Sequence turnoff (MS\--TO) of the old population present in the bulge for most of the fields.
In principle, such photometry should enable the study of the stellar ages, which are still one of the most controversial open question \citep[see][and references therein]{surot+18}.
Unfortunately, \citet{alonso_vvv} photometry based on DoPhot does not include an evaluation of completeness on every field, but instead it relies on three \textit{representative} fields. This is insufficient for studies focusing on stellar dating of the bulge population or those for which having homogeneous field\--to\--field stellar counts are mandatory.

To overcome this limitation we have carried out a new PSF\--fitting photometry using DAOPHOT, which allows us to make use of extensive artificial stars tests and error analysis simulations on all VVV bulge fields. 
The photometric data reduction and the simulations of the bulge VVV dataset necessarily require very long CPU time, therefore we decided to limit our study only to 2 filters, J and K$_s$.
 
This paper presents the new photometric database of 196 VVV bulge fields that have enabled an accurate study of the bulge stellar ages \citep{surot+18}, as well as to derive a high spatial resolution \citep[to better than 1\,arcmin, which is the limit for previous maps,][]{oscar18} extinction map of the bulge \citep{surot+19}.

\section{The dataset}
\label{sec:dataset}

We use a combination of VVV J and $\mathrm{K_s}$ observations of bulge fields collected with the wide field near\--IR imager VIRCAM mounted at the VISTA 4\--m telescope at the ESO Paranal Observatory.  

VIRCAM is equipped with a mosaic of 16 Raytheon VIRGO 2048$\times$2048 detectors with gaps between the detectors of about 90\% of the chip size along the X--direction, and 42.5\% along Y.
The average pixel scale of the detectors is $0.339''$, with percent\--level variations across the whole detectors ensemble, resulting in each detector covering $\sim$\,133\,arcmin$^2$ on the sky. 
A single VIRCAM frame is called {\it pawprint}, and it consists of 16 single\--detector images (SDIs).

\begin{figure}
	\centering
	\includegraphics[width=9cm]{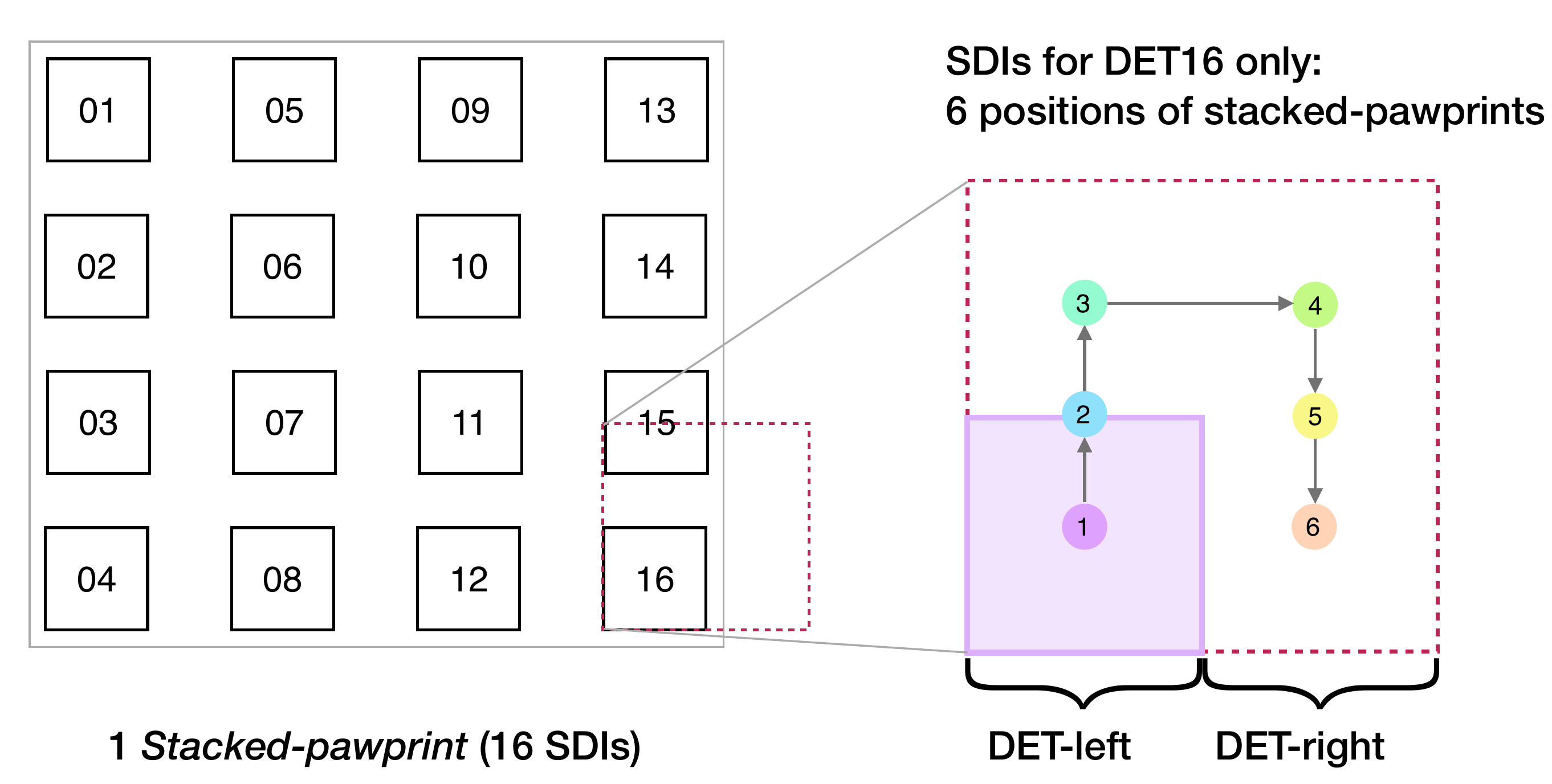}	\caption{{\it Left panel}: Representation of a stacked pawprint (black solid line). Numbers refer to the detector identifications, while the dashed red square defines the total area mapped by detector\,16 (DET16) when the 6 stacked pawprints are merged together. {\it Right panel:} Schematic pattern of DET16 SDI in the 6 stacked pawprints. Colors and numbers denote the order and position in which the 6 exposures are taken. The dotted red square refers to the total field of view covered by the mosaic of the 6 SDIs, while DET\--left and DET\--right define the total area mapped by positions $1+2+3$, and $4+5+6$, respectively.}
	\label{fig:pawpattern}
\end{figure}

The VVV observing strategy was designed to obtain a pair of pawprints jittered by $\sim$\,20$''$ to account for detectors bad cosmetics, at 6 different positions. 
The combination of the paired jittered pawprints is referred to as stacked pawprint (see left panel of Fig.\ref{fig:pawpattern}).

The offsets pattern between the 6 positions was properly defined in order to get a nearly homogeneous sky coverage of $\sim$\,$1.5\times1.2$\,deg, the so\--called tile. 
The right panel of Fig.~\ref{fig:pawpattern} shows a schematic view of the offsets pattern for any given SDI of the stacked pawprint.
In summary, a single tile is composed of 16$\times$6 SDIs (i.e. a stacked pawprint $\times$6 positions), per epoch, per filter. 

The exposure time per pawprint and epoch was only 4\,sec for $\Ks$ and $2\times6$\,sec for $\J$. 
With this strategy almost every pixel within a tile gets exposed at least twice, yielding effective exposure time of 8\,sec for $\Ks$ and 24\,sec for $\J$\--band for the stacked pawprints.
However, the overlap areas between stacked pawprints and edges of the tiles had 2\--6 times higher exposures causing the noise distribution within a tile to vary strongly with position in the sky. 
For this reason we decided to work on the stacked pawprint images (i.e. average of the two jittered exposures at each pawprint position), rather than using the final tile images.
Such images are available for download at Cambridge Astronomy Survey Unit (CASU\footnote{http://casu.ast.cam.ac.uk/}), after the corresponding raw science and calibration frames are processed by the VISTA data flow system pipeline \citep[v1.3][]{CasuPipe}.
For a more detailed description see \cite{saitovvv}.

\begin{figure}
	\centering
	\includegraphics[width=9 cm]{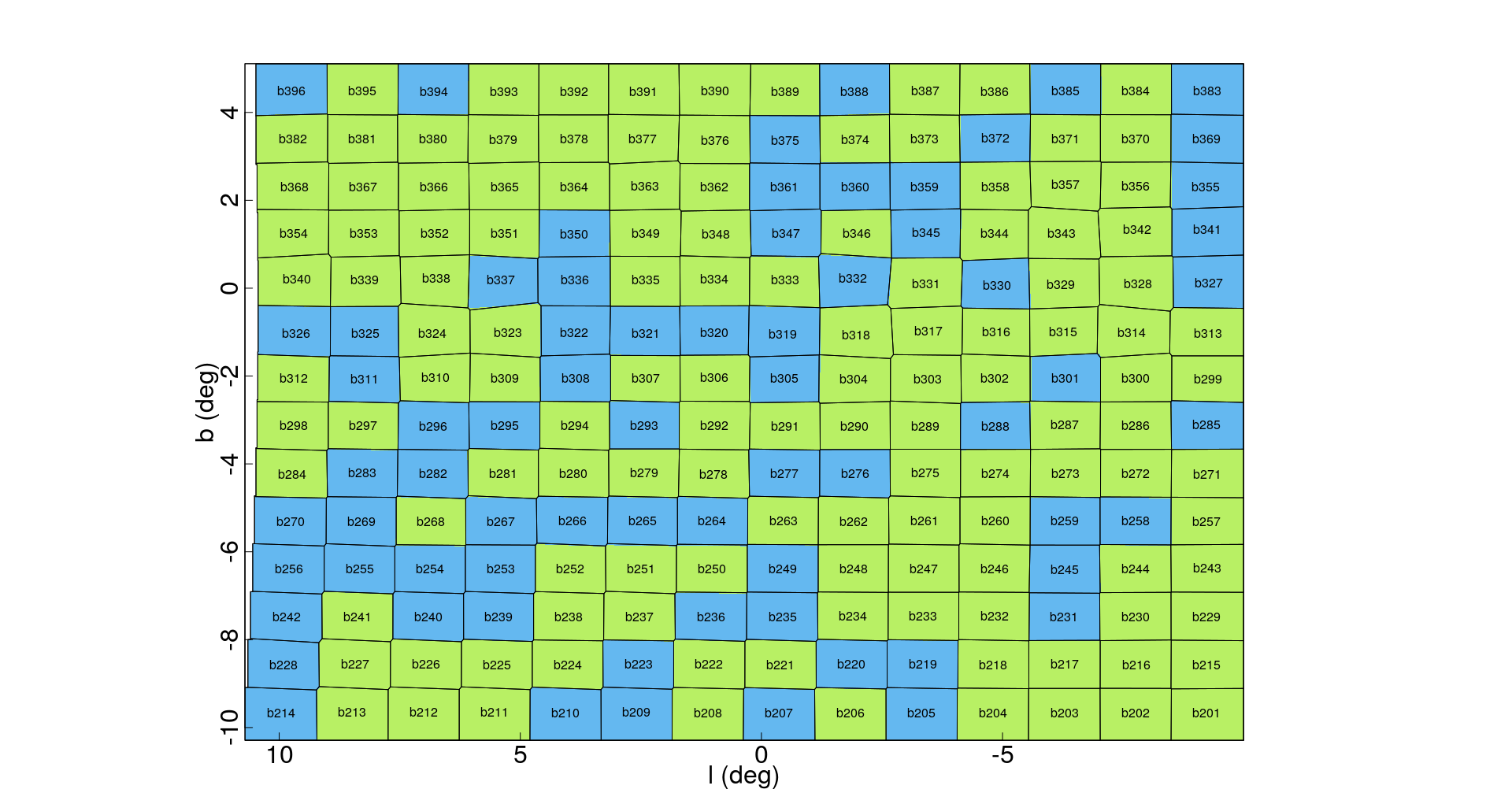}
	\caption{VVV survey bulge area and tile numbering. The color code refers to the number of epochs used to construct the photometric catalog of each tile: green for 2 epochs and blue for only 1 epoch.}
	\label{fig:epochmap}
\end{figure}

Figure~\ref{fig:epochmap} shows schematically the bulge area covered by the observations, and the official VVV tiles numbering used also in this work. 
Tiles for which 2 epochs in $\J$ and $\Ks$ have been used are highlighted in green, whereas those in blue have been obtained by using 1 epoch only.
In principle, we could have used 2 epochs for all 196 tiles because they have been observed twice in $\J$\--band and over 290 times in  $\Ks$. 
However, to obtain CMDs as deep and accurate as possible all images in both bands must have similarly good image quality (IQ). 
Therefore, for all tiles we first selected the $\J$ and $\Ks$ best IQ epoch.
Then, we use a second epoch only in those cases for which the second epoch in $\J$ had IQ similar (i.e. within 0.2") to the second best IQ epoch in $\Ks$.
Only 65\% of the tiles satisfies this selection criterion. 
The average IQ of the selected images is 0.75$''\pm$0.1 and 0.54$''\pm$0.04\footnote{Close to the instrumental PSF size, thus the best IQ that VIRCAM can deliver.} for $\J$ and $\Ks$ bands, respectively.
The complete list of stacked pawprint images (see Tab.\,\ref{tab:allima}) used in this work amounts to 3912, which correspond to a total of 62592 SDIs ($3912 \times16=62592$ SDIs).

\section{Data reduction}
\label{sec:psfphot}

The bulge region studied here is characterized by a very large star density, especially close to the Galactic plane ($|b| < 3^\circ$, $\gtrsim$\,1500\--3000 stars per arcmin$^2$), therefore to obtain accurate stellar photometry we used PSF\--fitting algorithms on each SDI, independently.
In this regard, each SDI has its own PSF, and as such, it is expected to represent an independent photometric system (i.e. zero\--points).
To obtain the final photometric catalog of any given tile we must first calculate and apply an internal detector\--by\--detector photometric calibration, and then combine all the SDI catalogs together.
In addition, we need to assess the photometric and systematics errors affecting the derived magnitudes, as well the completeness level (i.e. fraction of observed to truly present/recovered stars per color\--magnitude bin).

To this end, we make use of an ad\--hoc customized pipeline based on DAOPHOT, ALLSTAR \citep{daophot}, and ALLFRAME \citep{allframe} to extract the magnitudes from the SDIs.
Later for quick image coordinates transformations and internal cross\--matching we use DAOMASTER \citep{daomaster}.

\subsection{Initial parameters}
\label{sec:inpars}

The first step toward the catalog creation, is to propose a set of initial parameters to the DAOPHOT routine. For this purpose we first set the gain\footnote{http://www.eso.org/observing/dfo/quality/VIRCAM/reports/\newline HEALTH/trend\_report\_GAIN\_AVG\_HC.html} and read\--out noise (RON\footnote{http://www.eso.org/observing/dfo/quality/VIRCAM/reports/\newline HEALTH/trend\_report\_READNOISE\_AVG\_HC.html}) levels, as published in the ESO Health Check monitor for the VIRCAM instrument. From this database, we have taken the closest values to the time of observation for each SDI.

\begin{figure}
	\centering
	\includegraphics[width=.5\hsize]{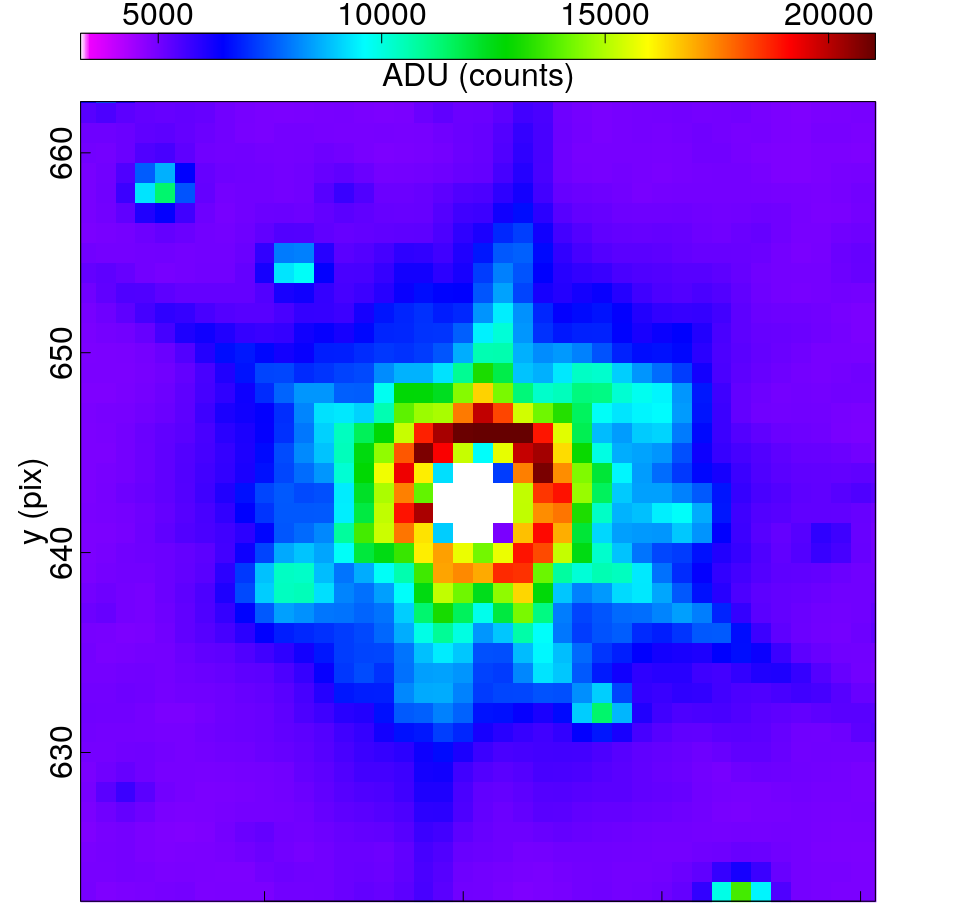}
	\includegraphics[width=.5\hsize]{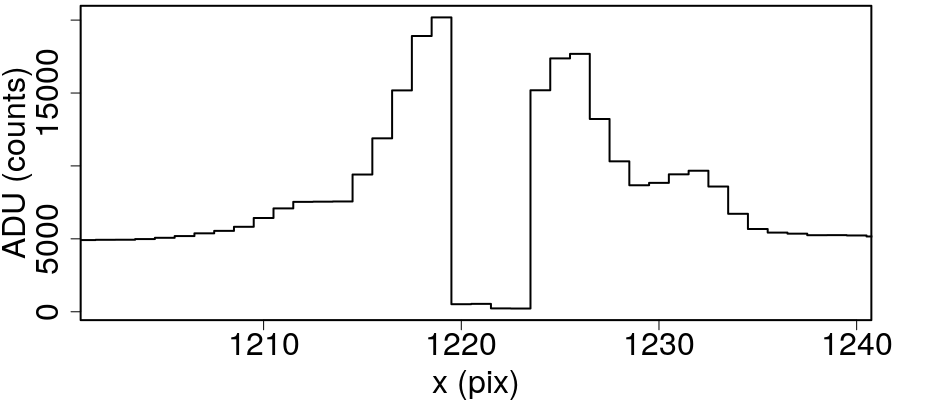}
	\caption[Zoom into a saturated star in the dataset, for detector\,05.]{Zoom in to a saturated star in detector\,05. The top panel shows a heatmap of the ADU counts, while the bottom panel is an horizontal cut through the center of the star. In cases like this very bright star the fault is evident and often leads to artifacts around it, but for stars moderately within the non\--linearity and saturation regimes, the central dip is much more subtle and cannot be filtered out simply.}
	\label{fig:saturated}
\end{figure}

From the VIRCAM user manual\footnote{http://www.eso.org/sci/facilities/paranal/instruments/vircam/doc.html} we obtain for each detector the recommended analog\--to\--digital unit (ADU) value corresponding to the linearity regime. 
However, we found that the values listed in the manual not always reflect what is observed in the SDIs, possibly because the reduction process due to dark correction, flat\--fielding, read\--out mode and combination of the jittered pair frames, has slightly changed the baseline counts for each FITS. 
In practice, this means that the ADU levels in the images never reach the saturation threshold in the user manual. Furthermore, due to the double correlated read mode, very bright stars have a {\it hole} with near\--zero counts in their center, surrounded by a somewhat smooth ring (see Fig.~\ref{fig:saturated}). In addition, non\--linearity deviations in the few\--percent range are already present at the 10,000 ADU, but these are well known and handled by the CASU pipeline.

%


Using these \textit{saturated} stars as a guideline, we have defined our highest count limit to 18,500 counts, which is roughly the level at which we can still find star\--like (i.e. PSF) signals, minus some arbitrary conservative margin of 1,000 counts. 
In case of detector\,05, which has a notoriously lower listed high\--count limit in the instrument manual, we have decided to reduce the high\--count limit value to 15,500. We will henceforth refer to this as the \textit{saturation limit}, even though, as stated here, it is not directly to the point of actual saturation, but when the images start being indirectly affected by it. 

By taking the brightest stars in the images and measuring the approximate radius from their center at which the ADU rises well above the local sky level, we have arrived at 15 pixels to be a reasonable value for the PSF radius. 
This value has been adopted throughout the whole reduction procedure that leads to the catalog extraction, regardless of detector.

In order to determine the Full Width at Half Maximum (FWHM) of the stars we proceeded as follows: {\it i)} we select 2,000 stars by using the DAOPHOT/PICK subroutine (pick\--stars) from each image, {\it ii)} we estimate the FWHM from the profile based on the maximum number of counts and the surrounding background, and {\it iii)} we obtained a general FWHM for the image from the median of the ensemble. 
We adopted a PSF\--fitting radius equal to the FWHM.

\subsection{PSF calculation}
\label{sec:psfcalc}

To construct the PSF model for each SDI we first run the DAOPHOT/FIND and DAOPHOT/PHOT routines and selected 450 bright isolated stars that are still well below the saturation limit (see \S\ref{sec:inpars}) by using the DAOPHOT/PICK subroutine.

We then proceed to iteratively filter out stars with a bad profile shape and bad\--pixels, as defined by DAOPHOT, until no star has bad\--pixels in its fitting radius and the final sample has no profile outliers.

\begin{figure}[t]
\centering
\includegraphics[width=.3\hsize]{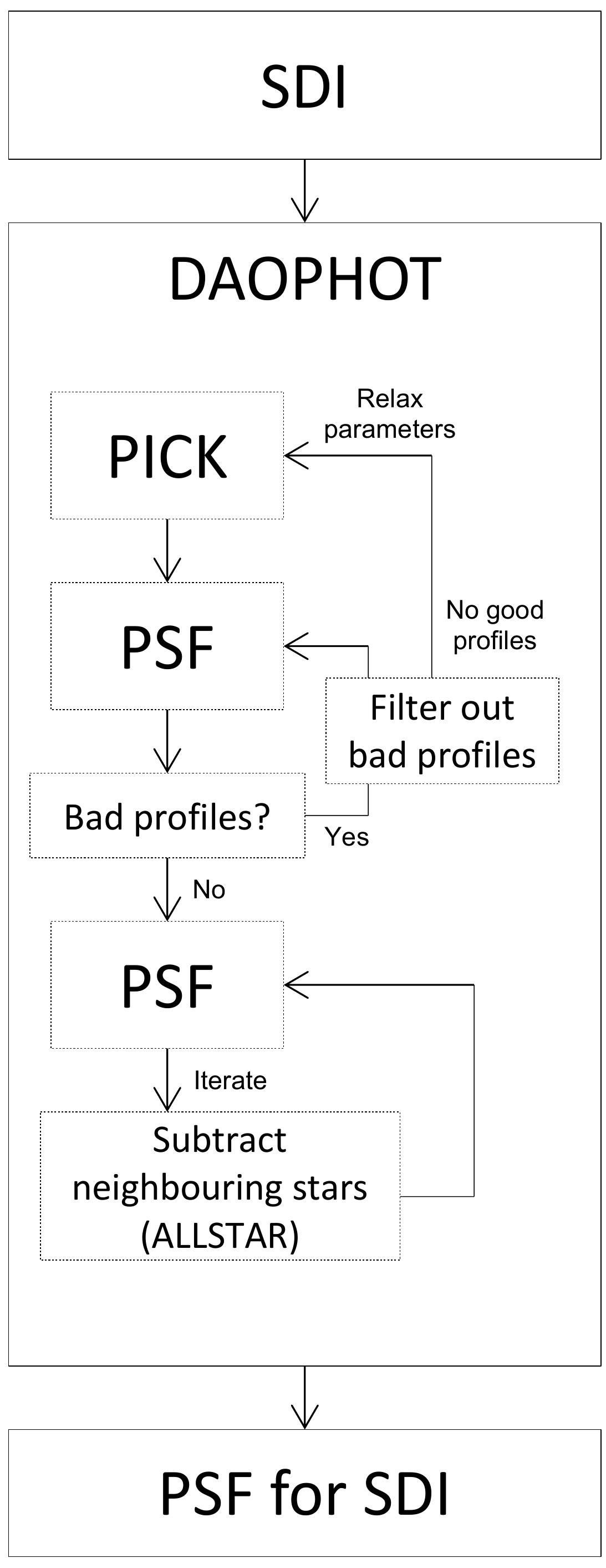}
\caption{Flow diagram of the computational routine used to obtain the PSF model of each SDI}
\label{fig:psf}
\end{figure}

Once the selection of PSF\--stars has been made, we calculate their profile using a Moffat distribution with $\beta = 2.5$ and allowing for a quadratic XY variation of the PSF across the image.
The derived PSF\--model is used as input parameter in the ALLSTAR routine, allowing to remove neighboring stars, and subsequently re\--fitting the PSF profile. 
An iteration of 5 times usually led to a convergent profile for most SDIs. 
However, for certain problematic cases, it was necessary to reduce the counts saturation limit (i.e. in steps of 1,000 ADU). 
The computational routine leading to the final PSF\--model is shown in Fig.~\ref{fig:psf}.

\subsection{SDIs mosaic and catalog calibration}
\label{sec:mosframes}

As mentioned in \S\ref{sec:dataset}, each VVV tile in each filter and epoch, is a mosaic of 6 stacked pawprints $\times16$ detectors, that is 96 SDIs, most of which have negligible spatial overlap.
For each VVV tile, all SDIs have been processed with DAOPHOT/ALLFRAME routine following the recommendations in the "Cooking with ALLFRAME v.3" document \citep{turner97} and the PSF\--model as derived in \S\ref{sec:psfcalc}.
However, because ALLFRAME improves the quality of the PSF fitting only for sets of images with some spatial overlap, we run it simultaneously on the set of 3 SDIs corresponding to the 3 vertical positions of the offsets pattern shown in Fig.\,\ref{fig:pawpattern}.
That is, we run ALLFRAME on SDI $1+2+3  (\times 2$ filters $\times$ N epochs) mosaicing into what we call DET\--left, and on SDI $4+5+6$, mapping into DET\--right.
The $16\times2$ DET\--$side$ $\J$ and $\Ks$ catalogs have been first combined into 32\,$\J\Ks$ catalogs, and then joined into a single one for each VVV field, which is what we provide to the community through the ESO Science Archive.

\begin{figure}
	\centering
	\includegraphics[width=\hsize]{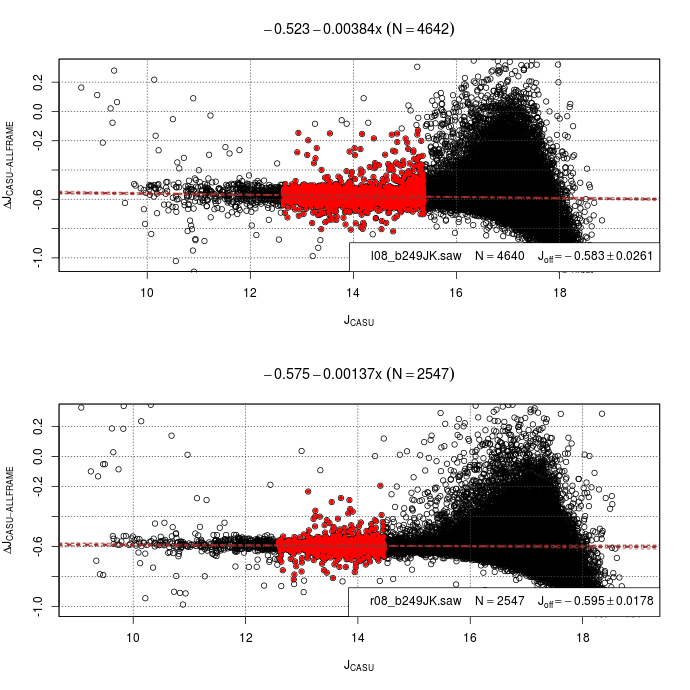}
	\caption[Example plot for the CASU\--ALLFRAME calibration for detector\,08 of tile b249.]{Example plot for the CASU\--ALLFRAME calibration for detector\,08 of tile b249. Top panel refers to the {\it left} side, while the bottom panel to the {\it right} side. Displayed is the magnitude difference in $\J$ between CASU and ALLFRAME catalogues vs. CASU reference. The black circles follow all the matches, while the red dots mark the selected magnitude range where the zero\--point is calculated. The title of each plot refers to the linear fit performed on the red dots, as well as the number of stars from which it is derived. Also shown in the legend is the source catalog, the estimated zero\--point and the corresponding uncertainty. The criterion to select the red dots is dependent on a window from which we have the minimum spread in the $\Delta m$ relation.}
	\label{fig:deltamagCASU}
\end{figure}

The absolute magnitude calibration of the photometric joint $\J\Ks$ catalogs is obtained through cross\--correlation with the catalogs produced by CASU\footnote{Using the provided fitsio\_cat\_list program by CASU at http://casu.ast.cam.ac.uk/surveys-projects/software-release} for the same set of images. 
Specifically, we first transform the XY position of the detected sources in these catalogs into the absolute system RA\--DEC by using the WCS recorded in each image header.
Then the match with the CASU catalogs is done with STILTS \citep{stilts}, using a RA\--DEC separation criterion with a tolerance of $0.5$\,arcsecs, roughly 1.5 pixels. 
Overall, the match presents a natural spread of about $\pm (0.02 \-- 0.03)$\,mag for $\J$ and usually a 50\% higher for $\Ks$ (see Fig.~\ref{fig:deltamagCASU}).  
It is worth mentioning that the photometric catalogs produced by CASU and adopted here as reference system for the absolute calibration might carry a zero-point calibration that is biased in denser fields. 
Such bias may rise from possible mis-matches and blends in the cross-correlation between 2MASS and VVV images performed by CASU to derive the final VVV photometric system. 
A thorough analysis performed on a chip-by-chip basis is not yet available, therefore it cannot be taken into account in the present work. 
On the other hand, should new zero-points become available in the future, in the context of the VVVX survey \citep{vvvx}, the calibration of the current photometric compilation can be easily improved a posteriori.

We also perform internal crosschecks within different pawprint images for which we have significant overlap. From these matches, the general result is a well centered dispersion of magnitude difference, $\Delta m$, around 0 with nominal spread within $0.5\sigma$, with an intrinsic scatter of $\pm 0.01$\,mag for $\J$ and, again, about 50\% higher for $\Ks$. 
Exceptions arise, however. For instance, the stars from detector \#16 show systematically redder colors in the upper third part of the detector ($Y \gtrsim 1400$), for all images. 
This is due to a known defect in the detector with filter\--dependent sensitivity.

\section{Completeness}
\label{sec:compl}
The photometric completeness of the newly derived catalogs is assessed through artificial star experiments.
The basic idea is to add to the SDIs synthetic stars with the observed PSF and $\J$ and $\Ks$ magnitude spanning the range observed in the data.
Then re\--process the entire dataset containing the artificial stars of known magnitudes by following the same procedure described in \S\ref{sec:psfphot}, but skipping the construction of the PSF model.
The fraction of recovered stars with respect to those added provides an estimate of the photometric completeness.

In practice, for each DET\--side $\J\Ks$ catalog, we first define a color\--magnitude mask covering almost the whole corresponding CMD.
This mask is used to construct a uniform 2D distribution in color\--magnitude space, from which we draw pairs that define the $\J$ and $\Ks$ magnitudes of the injected stars ($m^{in}$ atlas). 
This allows us to have completeness information virtually on all stars present in the observed CMD with no waste in CPU time.

The artificial stars are then spatially distributed (i.e. XY) around a grid properly customized to avoid artificially increasing the crowding \citep[see][]{zoccali+03}.
Specifically, we use a hexagonal grid with distance between nodes of 30 pixels ($\sim$\,2$\times$PSF radius).
This allowed us to inject  $\sim$\,11,000 stars per DET\--side each time.
We repeated the process 10 times, for a total of up to $\sim$\,120,000 artificial stars per DET\--side, hence $\sim$\,240,000 stars per detector.

These modified images are then processed in exactly the same way as described in \S\ref{sec:psfphot}, with the exception of both PSF\--models and coordinate transformations steps, which are recycled from the photometry process, rather than redefined. 
This produces, for each detector, 2 catalogs $\J\Ks$ (left and right).

We note that since the catalogs themselves have joint $\J\Ks$, that is, there is no measure in one filter without the other, the final product is a completeness value that is a function of both magnitudes: $p = p(\J,\Ks)$.

Finally, we cross\--correlated these catalogs to the injection ones using STILTS, with a separation criterion in XY of at most 1.5\,pixels (similar to the calibration run), providing us now with a recovered magnitude $m^{rec}$ match to the corresponding $m^{in}$.

We consider a star as {\it recovered} if $|m^{in}-m^{rec}| < 0.75$\,mag for each filter \citep[see][]{sollima+07}.
Thus, the final completeness value $p(\J,\Ks)$ is simply defined as the ratio between the number of recovered stars and the number of injected stars per $(\J-\Ks) \-- \Ks$ bin, where the latter has been defined as  $0.14 \times 0.13\,\mathrm{mag}^2$.

We note that the completeness in general is different from one detector to another, with nominal $\pm 0.2$ variations around the $p = 0.5$ level, regardless of actual stellar density.

In addition to the $p(\J,\Ks)$ values, the completeness experiments provide an estimate of the total (systematic and photometric) uncertainty of the system. 
This estimate comes from a fourth\--degree polynomial fit to the dispersion of $m^{in}-m^{rec}$ per magnitude bin, and we will henceforth refer to it as the \textit{combined errors} $\Delta_\J$ and $\Delta_\Ks$.

Finally, in the available technical documentation at the CASU website\footnote{http://casu.ast.cam.ac.uk/surveys-projects/vista/technical/known-issues} several known issues are highlighted regarding the VISTA image quality. Most of them are either unavoidable or resolved by the time of the observations, but there are two precautions we thought would be best to take, and that was to not include detectors\,04 and 16 in the completeness analysis. 
In the case of detector\,04, the problem is mild and not always present, but we decided to exclude it regardless. For detector\,16, however, the defect is persistent and too hard to correct effectively. 

\section{Final photometric catalogs}
\label{sec:fincat}

The final product of the procedure described in the previous sections is a compilation of 196 photometric catalogs, one per each VVV bulge field (see Fig.\ref{fig:epochmap}) covering a total of $\sim$\,300\,deg$^2$ around the Galactic center.

For each star detected in a given tile the catalog provides: the equatorial and galactic coordinates [RA, DEC, l, b]; the magnitudes (in the VVV photometric system) with the corresponding photometric and  combined errors [$\J , \Ks, \sigma_\J, \sigma_\Ks, \Delta_\J, \Delta_\Ks$]; the completeness  [p($\J$,$\Ks$)]; the extinction [$\mathrm{E}(\J - \Ks)$, $\sigma_{\mathrm{E}(\J - \Ks)}$], and dereddened magnitudes [$\J_0$, $\Ks_0$] as newly derived in \citet{surot+19}; some parameters describing the quality of the PSF\--fit (\textit{sharpness}, $\chi^2$); the number of times a particular star was detected (rep); and a binary (i.e. base 2) flag tracing the detector(s) of origin. he binary flag can be used to refine a posteriori the photometric zero\--points should a new calibration based on VVVX data demonstrate the need for zero\--point change (see \S\ref{sec:mosframes}).

\begin{figure}      
	\centering
	\includegraphics[width=.9\hsize]{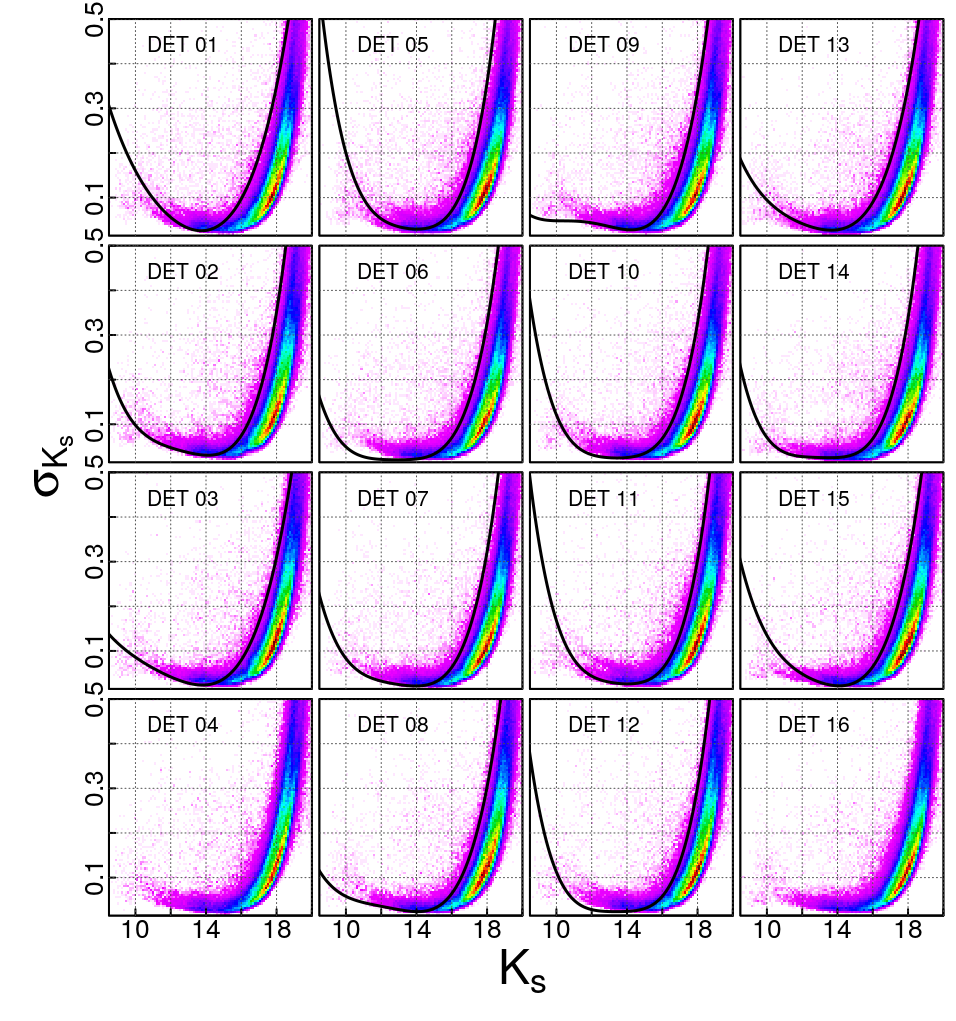}
	\caption{Photometric error profile for a sample field (b249). The 2D histogram in each panel displays the $\sigma_\Ks$ vs. $\Ks$ distribution of detected stars within a given detector (the detector number is labeled in each panel). Color coded for the density in each 2D histogram, from low (magenta) to very high (dark red) relative densities. The split or broadened sequences are due to the error mitigation in overlapping areas of a side ensemble. Detections in overlapping areas have smaller uncertainty. The black solid line refers to the {\it combined errors}: $\Delta_\Ks$ vs. $\Ks$, as calculated from the completeness experiments (except for detectors\,04 and 16, for which no completeness is available). }
	\label{fig:error}
\end{figure}

It is worth mentioning that to properly assess the photometric quality of each catalog one should not exclusively use the tabulated photometric errors (i.e. $\sigma_\J, \sigma_\Ks$), but rather the {\it combined errors} ($\Delta_\J, \Delta_\Ks$) described in \S\ref{sec:compl}. Indeed, as we show in Fig.~\ref{fig:error}, the combined effect of systematics and photometric uncertainties produces a spread in the recovered vs. injected magnitudes (black solid line) that is considerably larger than what one would expect from the photometric error alone (colored points) for stars outside the $16 \gtrsim \Ks \gtrsim 13$ region. 
In addition, Fig.~\ref{fig:error} evidences the known issue related to the saturation/non\--linearity of bright stars in the VVV $\mathrm{K_s}$\--band images ($\mathrm{K_s} \lesssim 12$). 
Of course this problem is also present in the J\--band, but the uncertainties up to $\mathrm{J} \sim 10$ are well in line with an exponential error profile, from both photometric and simulated sources.

The provided $\chi^2$ and \textit{sharpness} ($s$) values can be additionally used to flag and filter out poor and/or false detections from the catalog.
The $\chi^2$ refers to the quality of the star PSF\--fitting, and its value should be distributed around 1: i.e.  $\chi^2=1$ is a perfect fit, and any value far from 1 is a poor fit. 
The sharpness ($s$) provides a measurement of how round ($s=0$) the detection looks in the image. 

The photometric catalogs provided here include a photometric error cut so that stars in the catalogs have $sigma < 0.5$\,mag in both J and $\Ks$, but no cuts applied based on the quality of fit of the PSF, other than those inherent to the DAOPHOT/ALLFRAME extraction process.

The CMDs shown in Fig.~\ref{fig:cmds}, however, do include a number of quality filters, described in \citet{surot+18}, aimed at rejecting any detections that are unlikely to be real stars.
Those selections are conservative enough so that they do not affect the catalog completeness provided here. 
One should keep in mind that stronger selections that might produce thinner/better\--defined sequences on the CMD might change significantly the completeness and therefore should be applied with caution when studying star counts.
\section{Derived Color\--Magnitude diagrams}
\label{sec:cmds}

The entire photometric dataset is very extensive and diverse, depending mainly on the stellar density and extinction of each field.
Therefore, we show here the derived CMDs and corresponding average completeness maps only for a sample of eight tiles sparsely distributed across the bulge area (see Table~\ref{tab:cmds}).
The fields have been selected such that, through the relative comparison of their CMDs, one can easily and quickly appreciate how the quality of the photometry varies along the bulge minor axis or at large longitudes.

The CMD of each tile includes routinely between 1 and 5 millions stars, therefore in order to avoid saturation in the point\--scatter plots, in this work we always use Hess diagrams.
Figure~\ref{fig:cmds} shows the observed CMDs and completeness maps of the selected fields.

\begin{table}  
	\caption{Position and average color excess of the tiles sample for which we show the derived CMDs. The total number of detected sources in each tile is also given.}             
	\label{tab:cmds}      
	\centering          
	\begin{tabular}{ c c c c l l l }     
		\hline\hline       
		Name & $(l, b)$ & $<E(\mathrm{J-K_s})>^a$ & Detected\\ 
		  &  &   & stars \\ 
		\hline
b208 & $( +1.11^\circ , -9.67^\circ )$ & $ 0.076 \pm 0.025 $ & 1,748,959 \\
b249 & $( -0.45^\circ , -6.39^\circ )$ & $ 0.163 \pm 0.053 $ & 2,704,933 \\
b272 & $( -7.79^\circ , -4.23^\circ )$ & $ 0.399 \pm 0.091 $ & 2,853,891 \\
b283 & $( +8.28^\circ , -4.21^\circ )$ & $ 0.250 \pm 0.047 $ & 3,196,013 \\
b292 & $( +0.97^\circ , -3.14^\circ )$ & $ 0.316 \pm 0.072 $ & 4,133,744 \\
b333 & $( -0.49^\circ , +0.16^\circ )$ & $ 3.515 \pm 0.857 $ & 2,850,439 \\
b376 & $( +0.98^\circ , +3.38^\circ )$ & $ 0.852 \pm 0.247 $ & 3,830,302 \\
b384 & $( -7.78^\circ , +4.51^\circ )$ & $ 0.262 \pm 0.066 $ & 3,363,181 \\
		\hline \hline
	\multicolumn{4}{l}{$^{a}$ Color excess from \cite{surot+19}, }\\
	 \multicolumn{4}{l}{average and standard deviation are taken over the whole tile.}
	\end{tabular}
\end{table}
\begin{figure*}
	\centering
	\includegraphics[width=0.4\hsize]{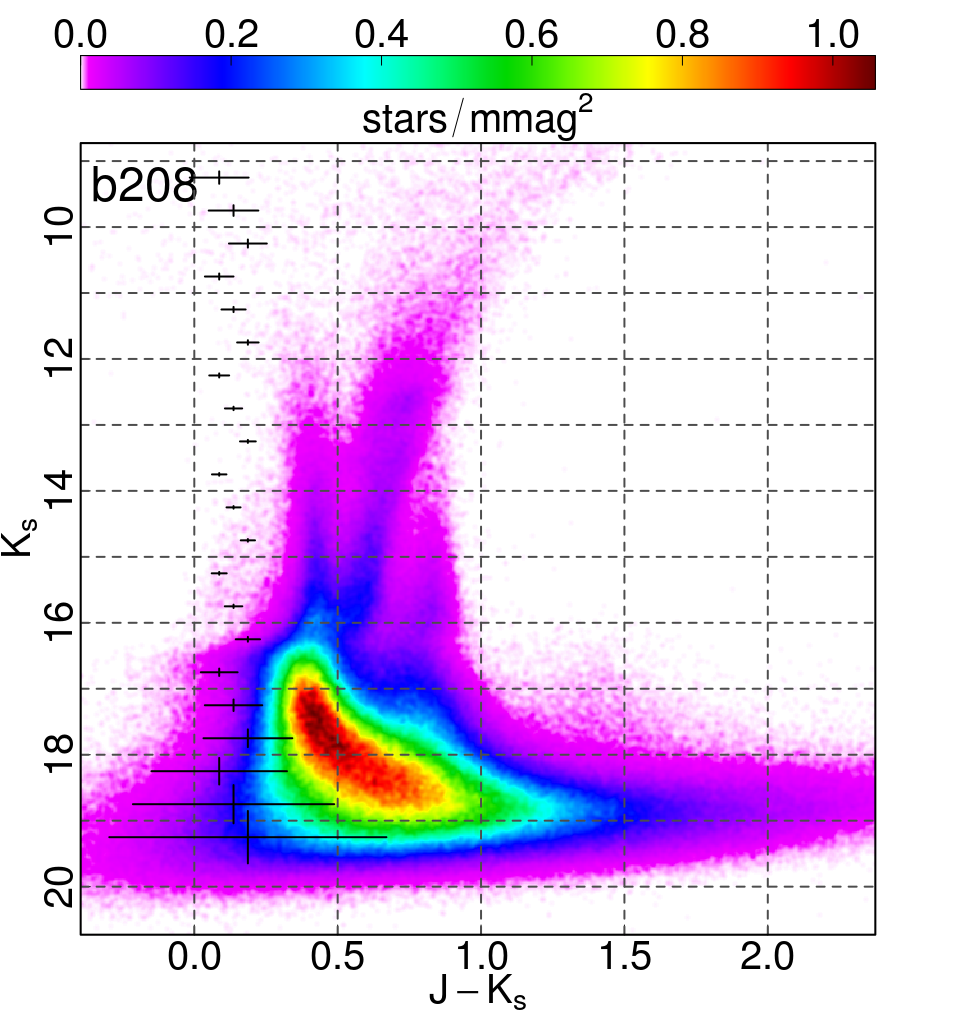}
	\includegraphics[width=0.4\hsize]{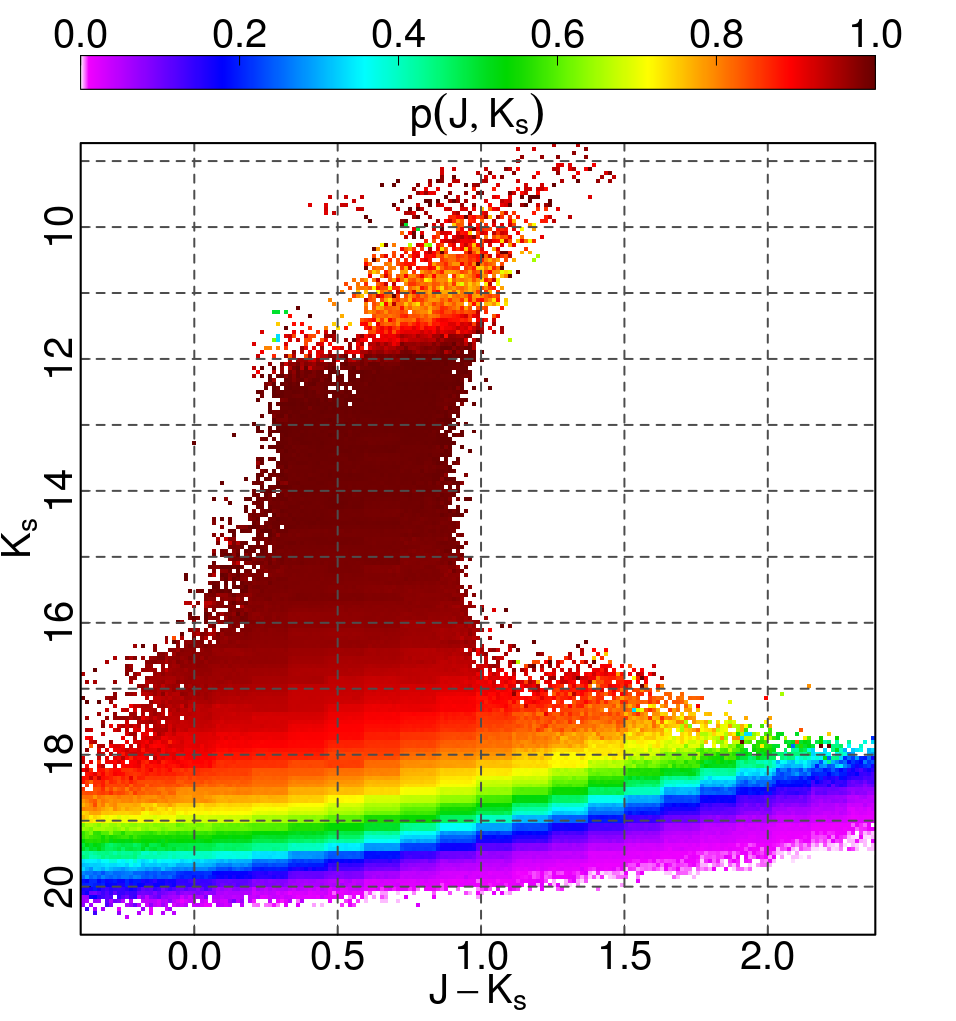}
	\includegraphics[width=0.4\hsize]{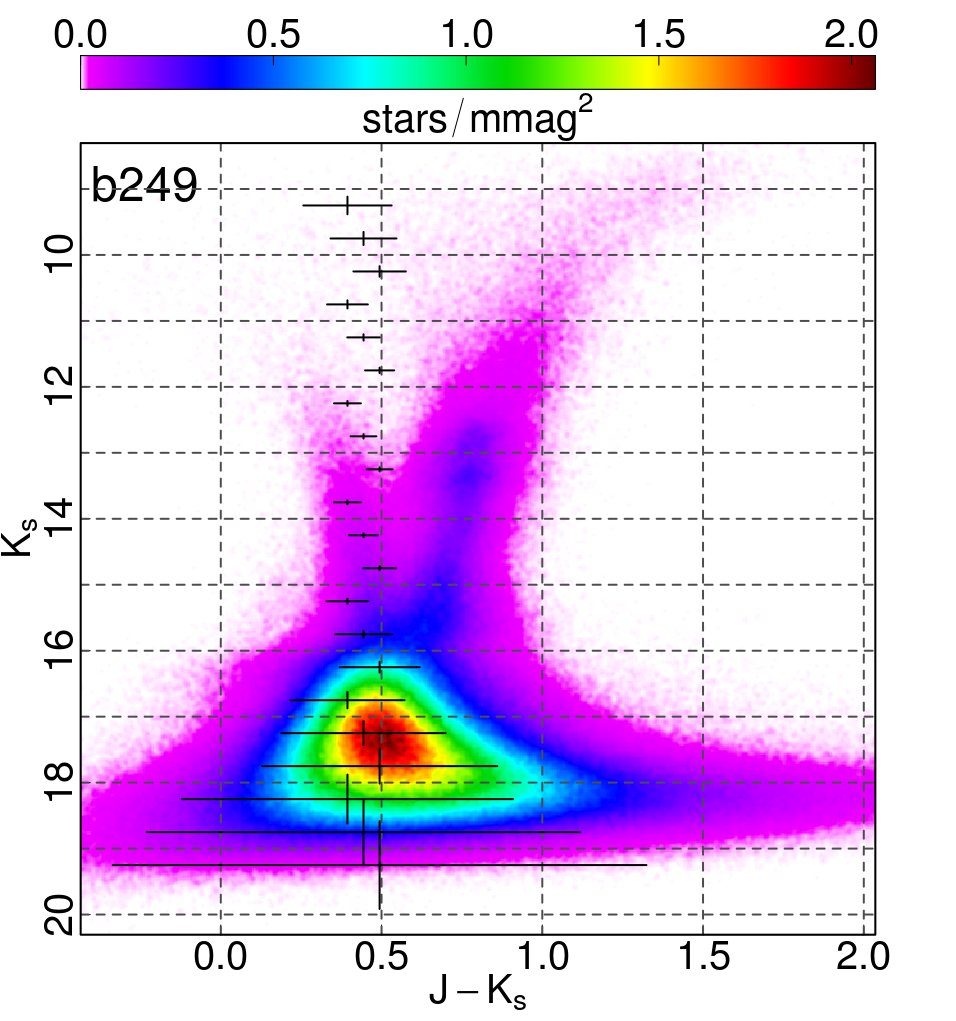}
	\includegraphics[width=0.4\hsize]{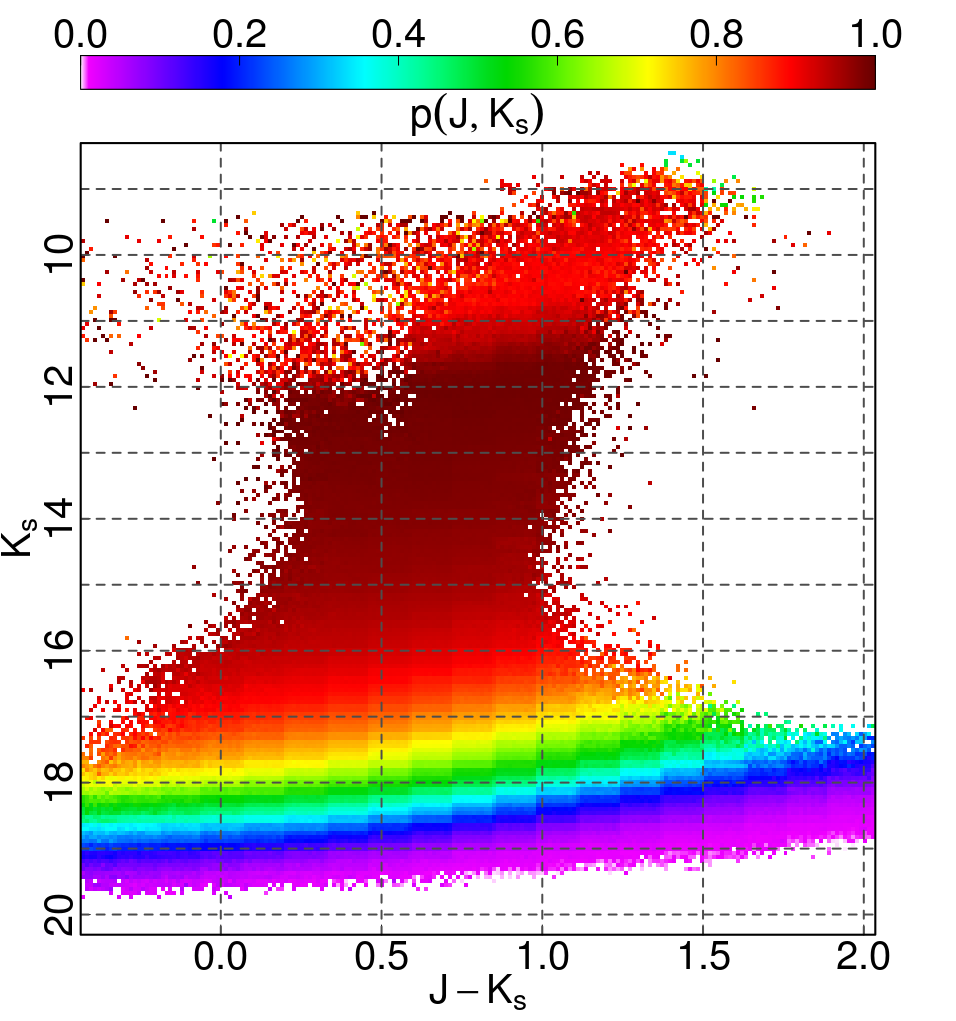}
	\includegraphics[width=0.4\hsize]{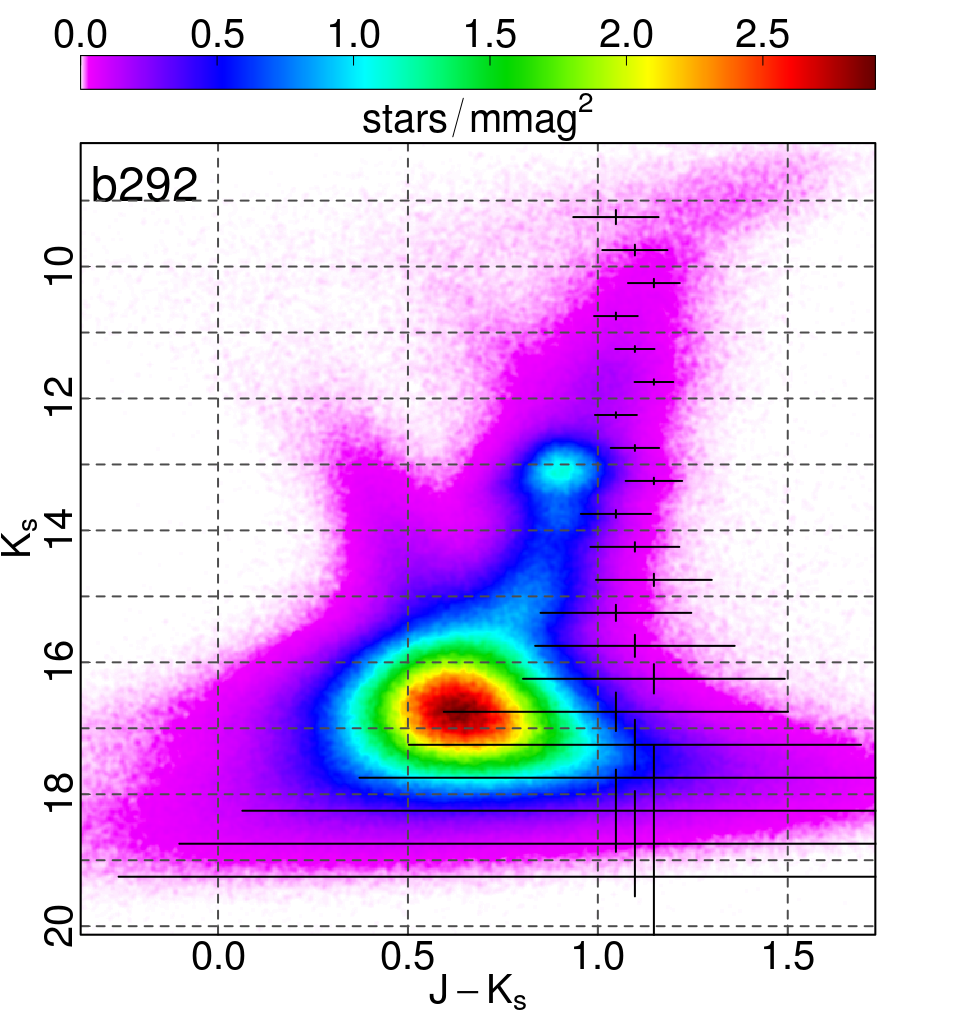}
	\includegraphics[width=0.4\hsize]{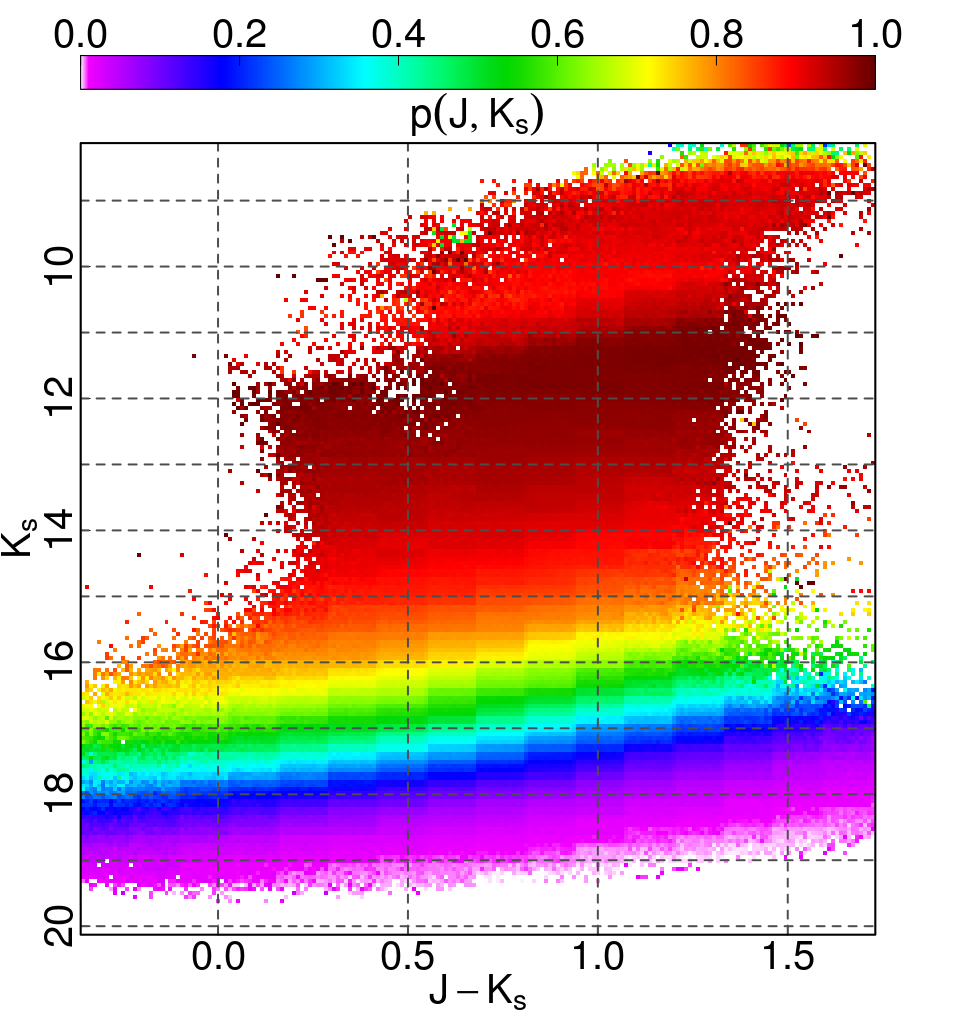}
	\caption{Hess diagram of the selected 8 tiles with typical color and magnitude errors shown as crosses at their respective $\Ks$ reference level (left panel), and corresponding photometric completeness map, averaged for all detectors except 04 and 16 (right panels). The VVV name of the field is labeled in each plot.}
	\label{fig:cmds}
\end{figure*}

\begin{figure*}
	\ContinuedFloat
	\centering
	\includegraphics[width=0.4\hsize]{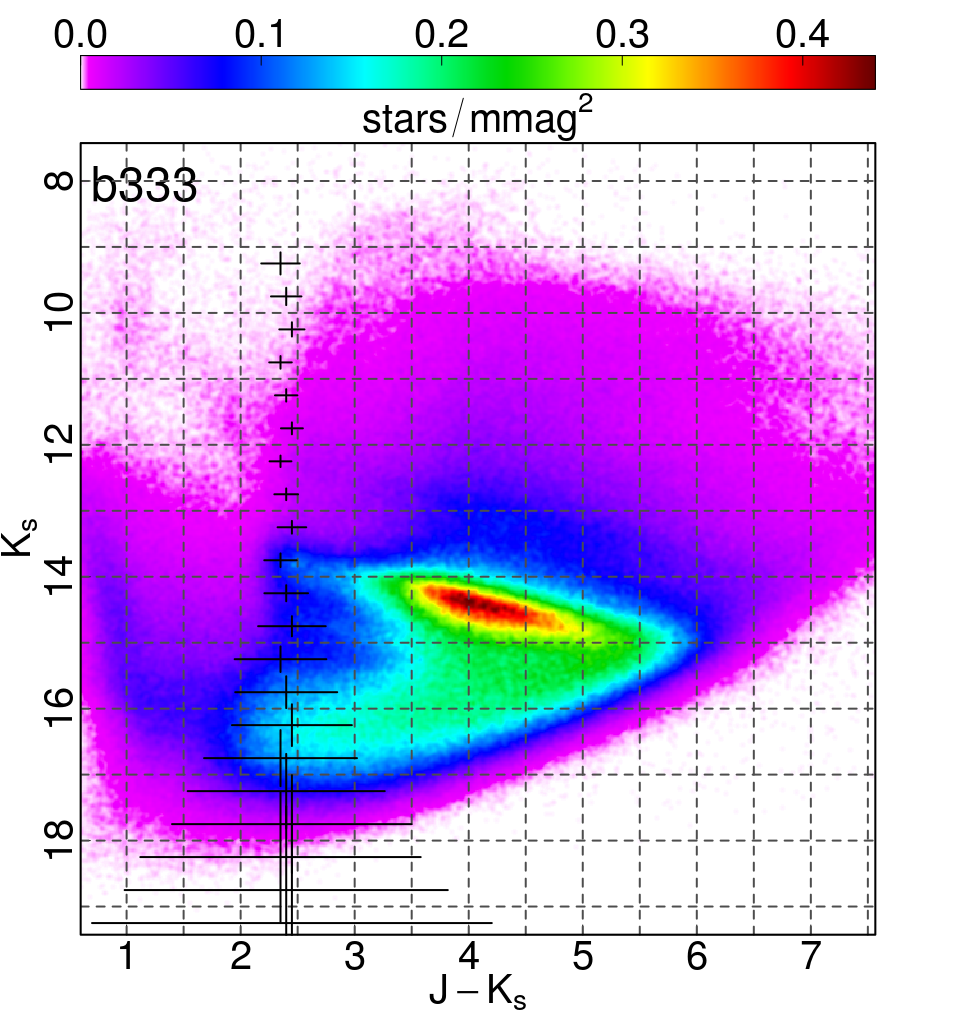}
	\includegraphics[width=0.4\hsize]{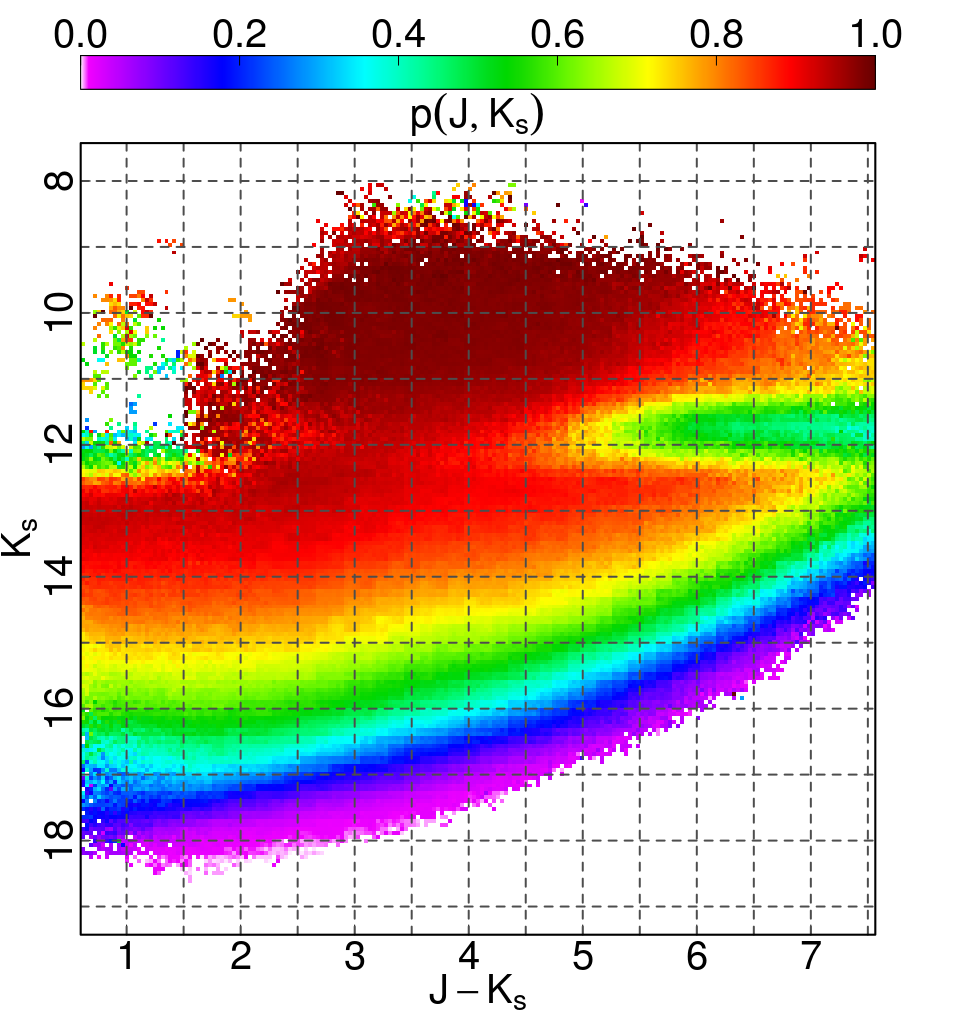}
	\includegraphics[width=0.4\hsize]{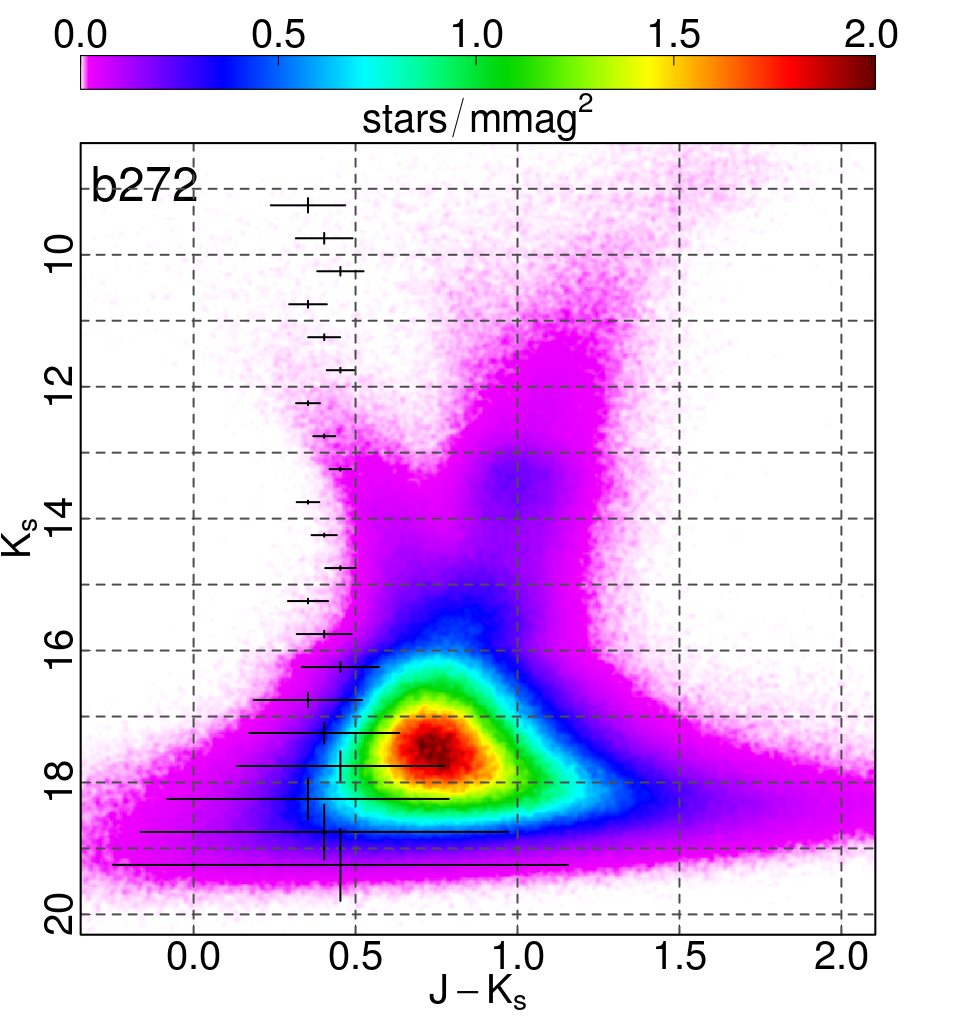}
	\includegraphics[width=0.4\hsize]{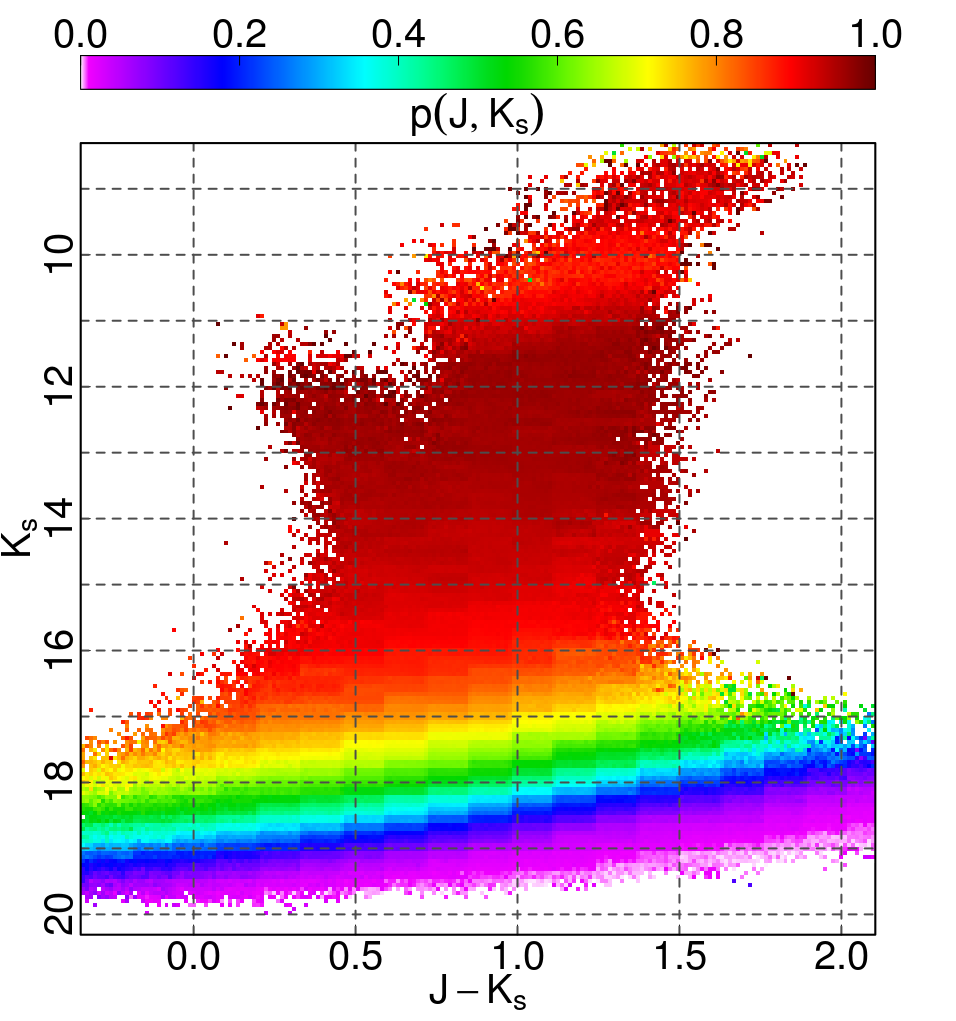}
	\includegraphics[width=0.4\hsize]{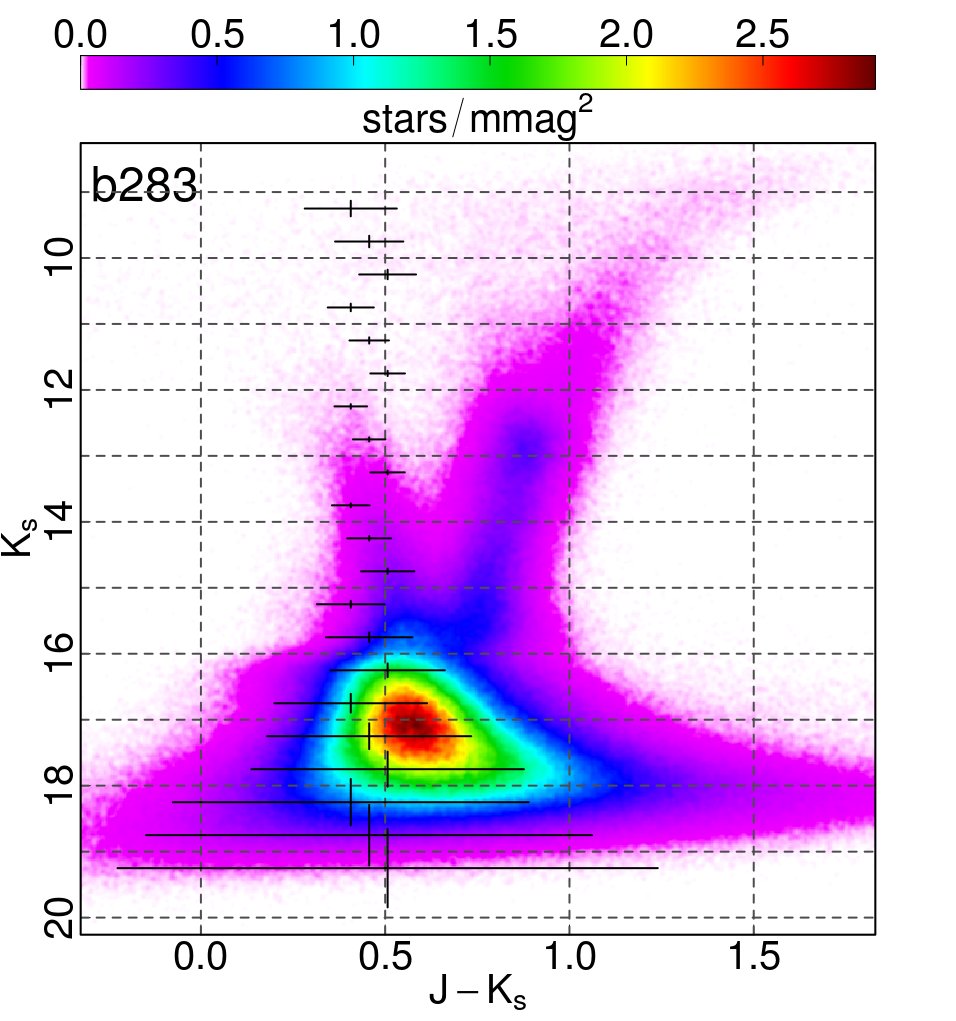}
	\includegraphics[width=0.4\hsize]{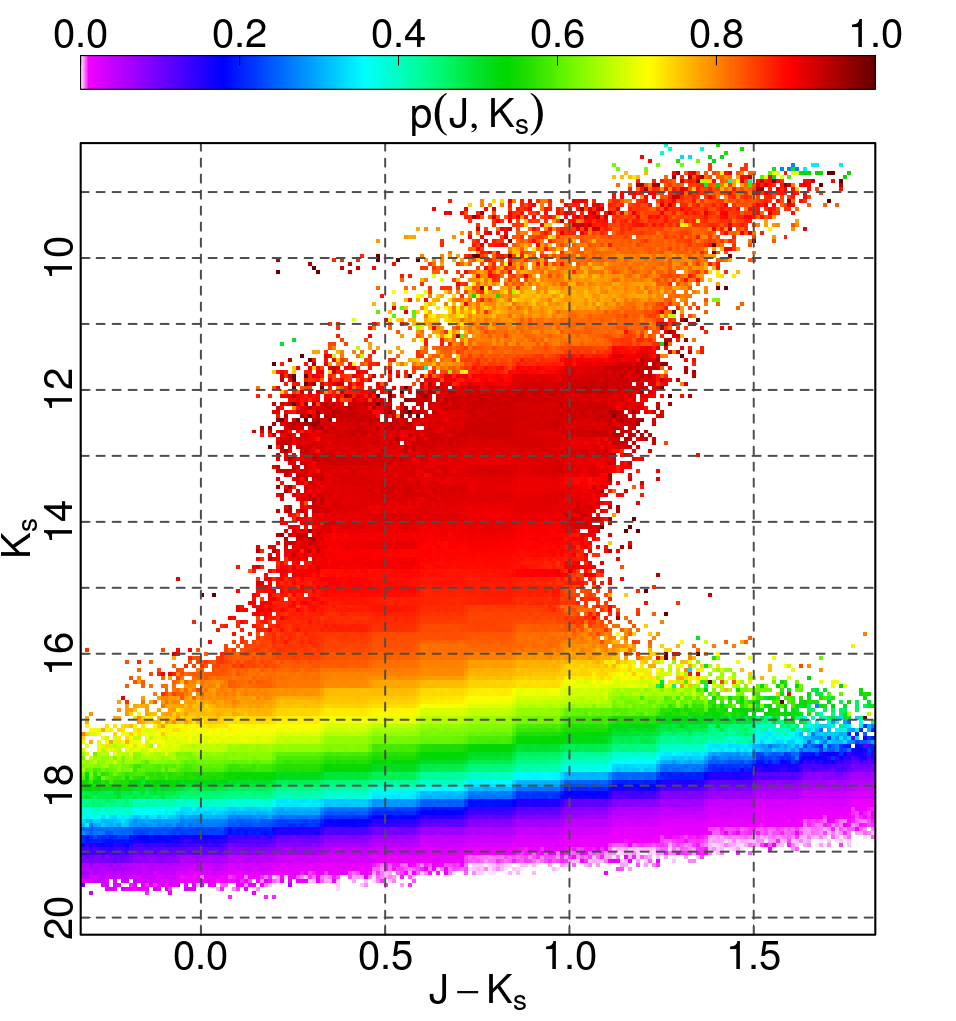}
	\caption{Continued.}
	\label{fig:cmds_2}
\end{figure*}

\begin{figure*}
	\ContinuedFloat
	\centering
	\includegraphics[width=0.4\hsize]{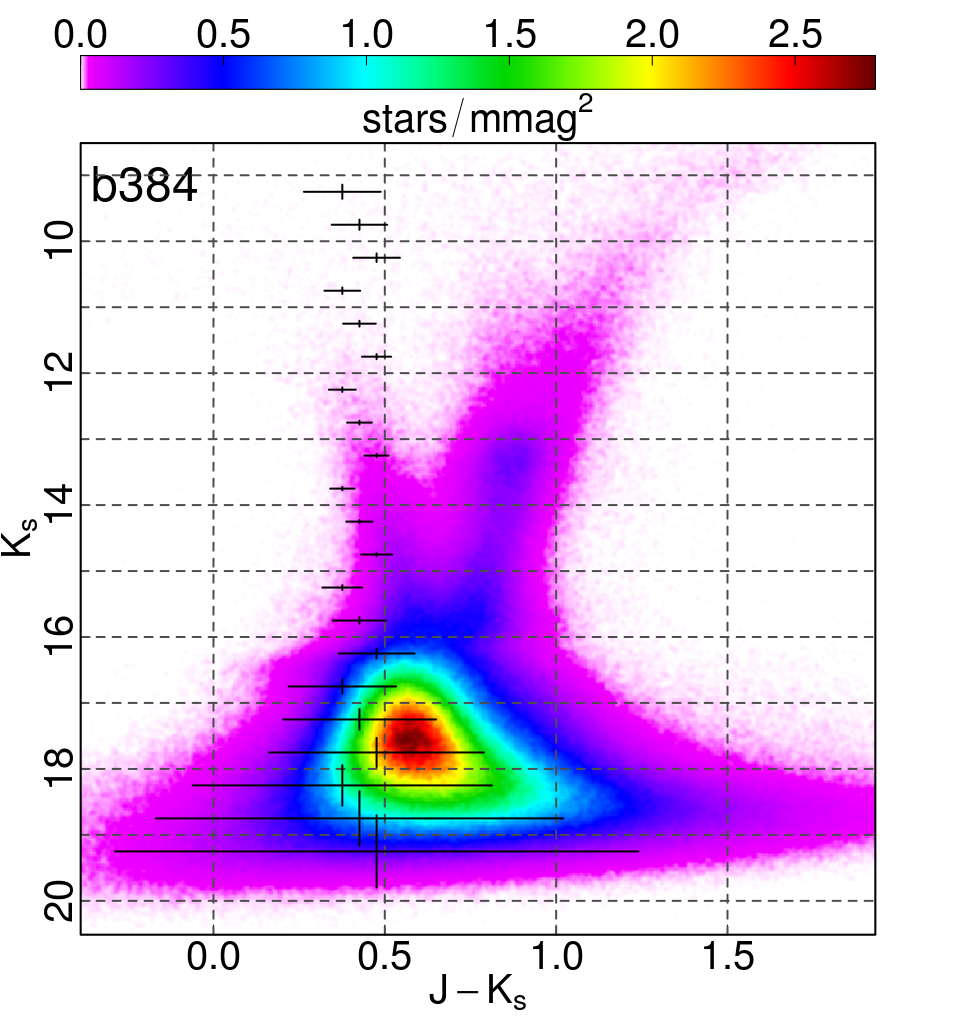}
	\includegraphics[width=0.4\hsize]{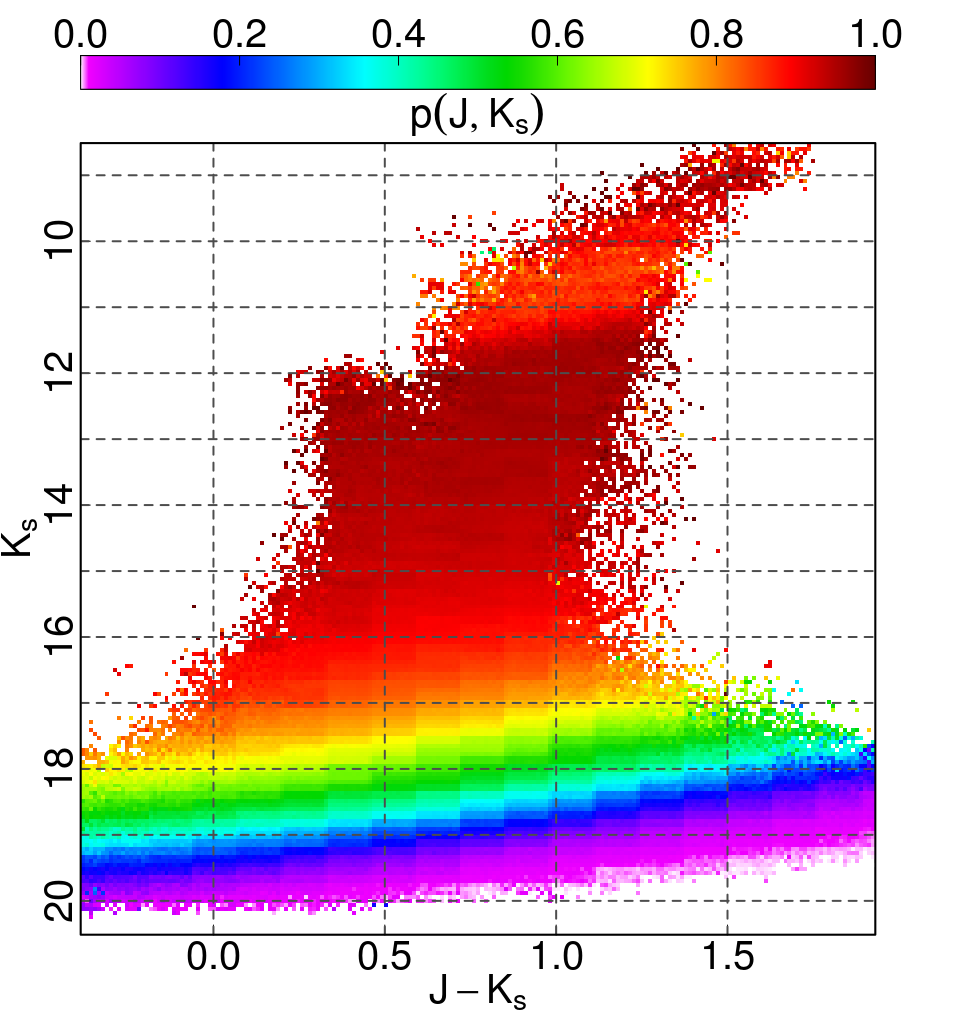}
	\includegraphics[width=0.4\hsize]{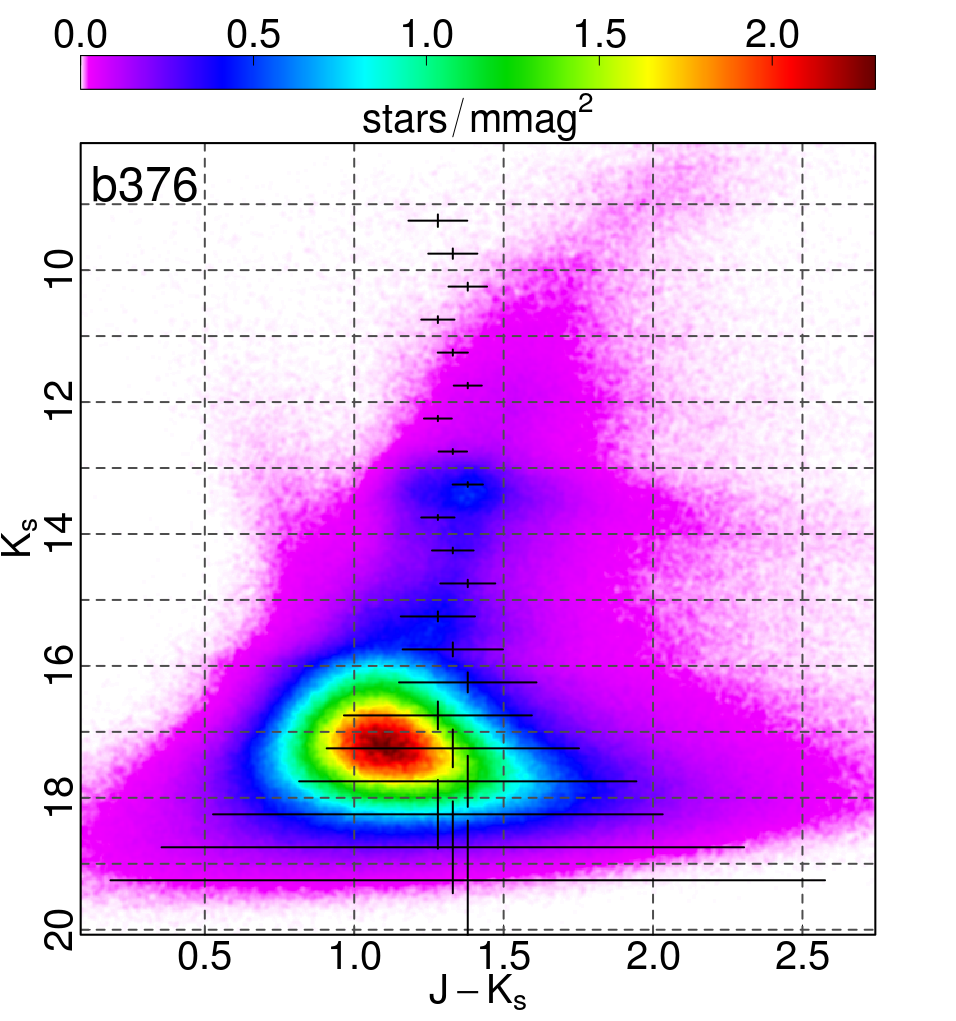}
	\includegraphics[width=0.4\hsize]{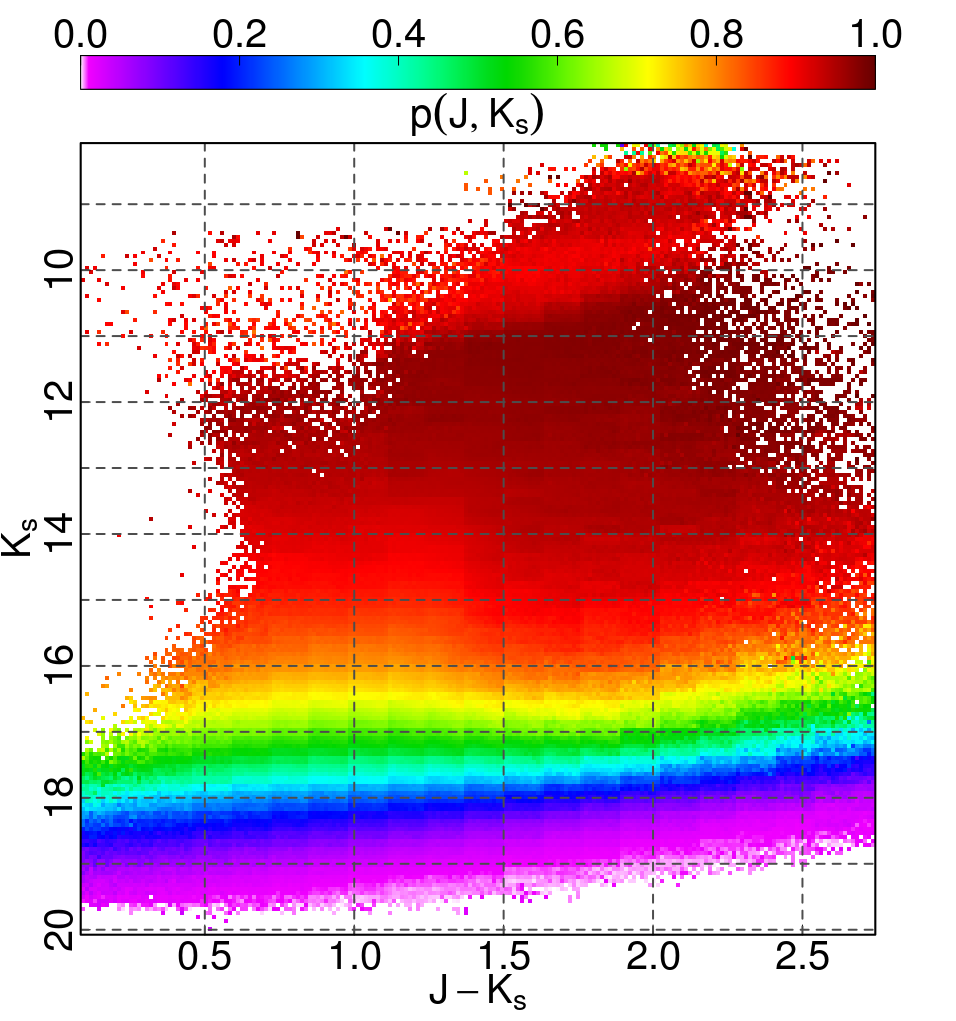}
        \caption{Continued.}
	\label{fig:cmds_3}
\end{figure*}  


With the exception of the tile b333, the common features of all CMDs are: the well defined bulge red giant branch (RGB\-- reddest vertical sequence $\Ks\gtrsim16$), the bright portion of the main sequence (MS) of the disk (bluest vertical sequence  $\Ks\gtrsim16$), the bulge RC (stellar overdensity along the brightest portion of the bulge RGB and generally between $14 > \Ks > 12$), the brightest end of the evolved disk population (vertical plume departing from the RGB towards the blue in the CMD at $\Ks\lesssim11$, but merging with the bulge MS below that limit), and the bulge MS that overlaps with the faint portion of the disk MS. Also noteworthy are the incomplete upper RGB and the mostly missing AGB due to the saturation and non\--linearity limitations.

The color spread of all sequences in the bright part of the CMD ($\Ks\lesssim16$) is mostly  caused by the differential reddening, metallicity spread and distance depth, whereas at faint magnitudes the photometric and systematic errors become predominant, smearing out the color of MS stars over a broad magnitude range.
The photometric limit is quite constant ($\Ks\sim19.5$) within the sample tiles, however as expected because of the different crowding and extinction, the photometric completeness varies substantially between the fields. For $\Ks\lesssim12$ there is a sizable dip in completeness, with a sharp drop to zero for $\Ks\lesssim9$.
The bulge MS\--TO is close to the {\it hot spot}, i.e. most dense area in the Hess diagrams, however due to the superposition of young disk MS and the increasing dispersion, it is impossible to ascertain this evolutionary phase easily. 
That said, in \citet{surot+18} we successfully demonstrated that the determination of the stellar age by means of the MS\--TO luminosity is feasible, providing that enough attention is paid to the observational effects.

The fields b208, b249 and b333 deserve special attention because, given their location, the derived CMDs show peculiar features.
Specifically, in the CMD of b208 one can identify the local K and M dwarf sequence as the vertical plume ( $\Ks\gtrsim16, (\J-\Ks)\sim 0.8)$, redder than the bright disk MS and bulge RGB. 
This sequence is usually only evident in the outskirts of the bulge area, at very large heights from the plane \citep[i.e. $b\lesssim-8^\circ$, see also][]{saitodemog}.
Moving towards the center, because of increasing reddening and stellar density, the local dwarf sequence progressively disappears because it is {\it swallowed} by the bulge RGB.

Tile b249 is particularly interesting because is located in the region where the X\--shape of the bulge is evident. 
Indeed, from the derived CMD one can observe the presence of 2 well separated RC (i.e. two apparent overdensities along the RGB at $\Ks\sim12.9$ and $\Ks\sim13.2$), which are the signatures of two southern arms of the X\--shape structure crossing the line\--of\--sight \citep{mcwzoc10,nataf+10,saito+11,wegg+13,ness+16}.

Tile b333 is also peculiar because it is located on the Galactic center region, and as such it is the most heavily reddened field in our sample \citep[i.e. $0.7 \lesssim \mathrm{A}_\Ks \lesssim 2.4$, ][]{surot+19}.
The derived CMD is the shallowest in the sample, and barely covers the entire RC population with a 50\% completeness level.
We note that the peculiar completeness trend, seen in Fig.~\ref{fig:cmds}, for b333 is due to a combination of many factors such as {\it i)} the average per color\--magnitude bin in this summary plot; {\it ii)} the differences in completeness between detectors; {\it iii)} the high differential and absolute extinction; and {\it iv)} the rapidly declining  $\J$ completeness with for red stars.
Finally, because the large extinction affects mostly only the bulge, the blue plume corresponding to the intervening evolved disk stars is clearly traceable in the CMD (see vertical sequence in the range $1\lesssim(\J-\Ks)\lesssim2.5$ and $8\lesssim\Ks\lesssim12$).
In this respect, for studies of the evolved disk populations along the bulge line\--of\--sight, the innermost tiles affected by large reddening are ideal because the bulge\--disk (background\--foreground) separation is quite clear.

%
\section{Overview of the global photometry}
\label{sec:global}

Containing nearly 600 millions stars, the new photometric database represents one of the most accurate and complete census of the stellar population in the Milky Way bulge.

This photometric database represents a valuable asset for the whole community, enabling different studies ranging from accurate star counts to stellar age determinations. 
It is also very useful in the context of upcoming massive spectroscopic surveys (e.g., MOONS, 4MOST etc.) in the Milky Way by providing homogeneous catalogs for target selection over large area.
All catalogs are available for download at the ESO Science Archive\footnote{http://archive.eso.org/scienceportal/home}, along with a set of figures (i.e completeness map, stellar density map, Hess diagram) that describe the main properties of the entire photometric compilation, and provide a quick look at the global bulge morphology and stellar content.

\begin{figure*}[h]   
	\centering
	\includegraphics[width=9 cm, trim={7cm 0 6cm 0}, clip]{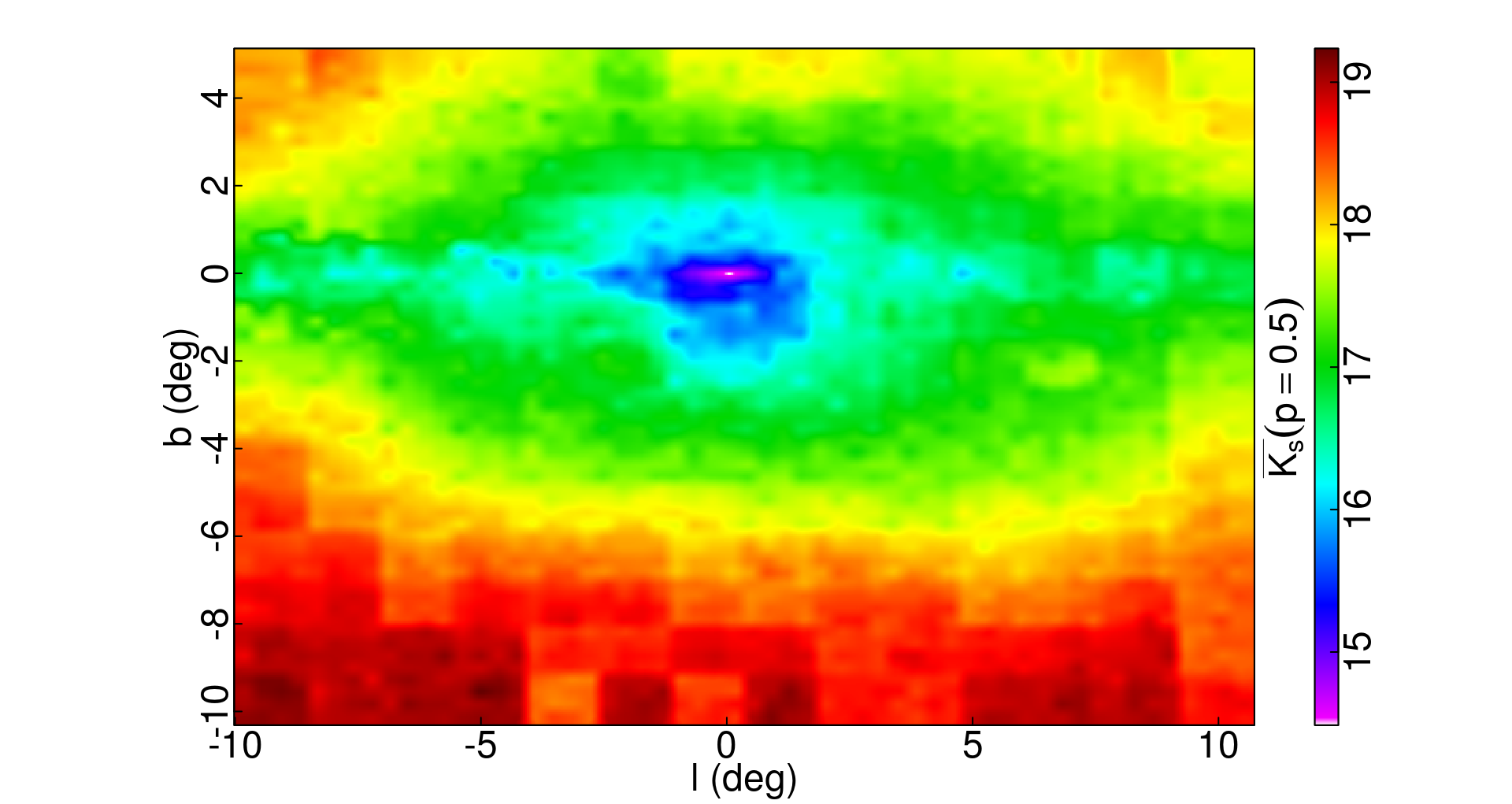}
	\includegraphics[width=9 cm, trim={7cm 0 6cm 0}, clip]{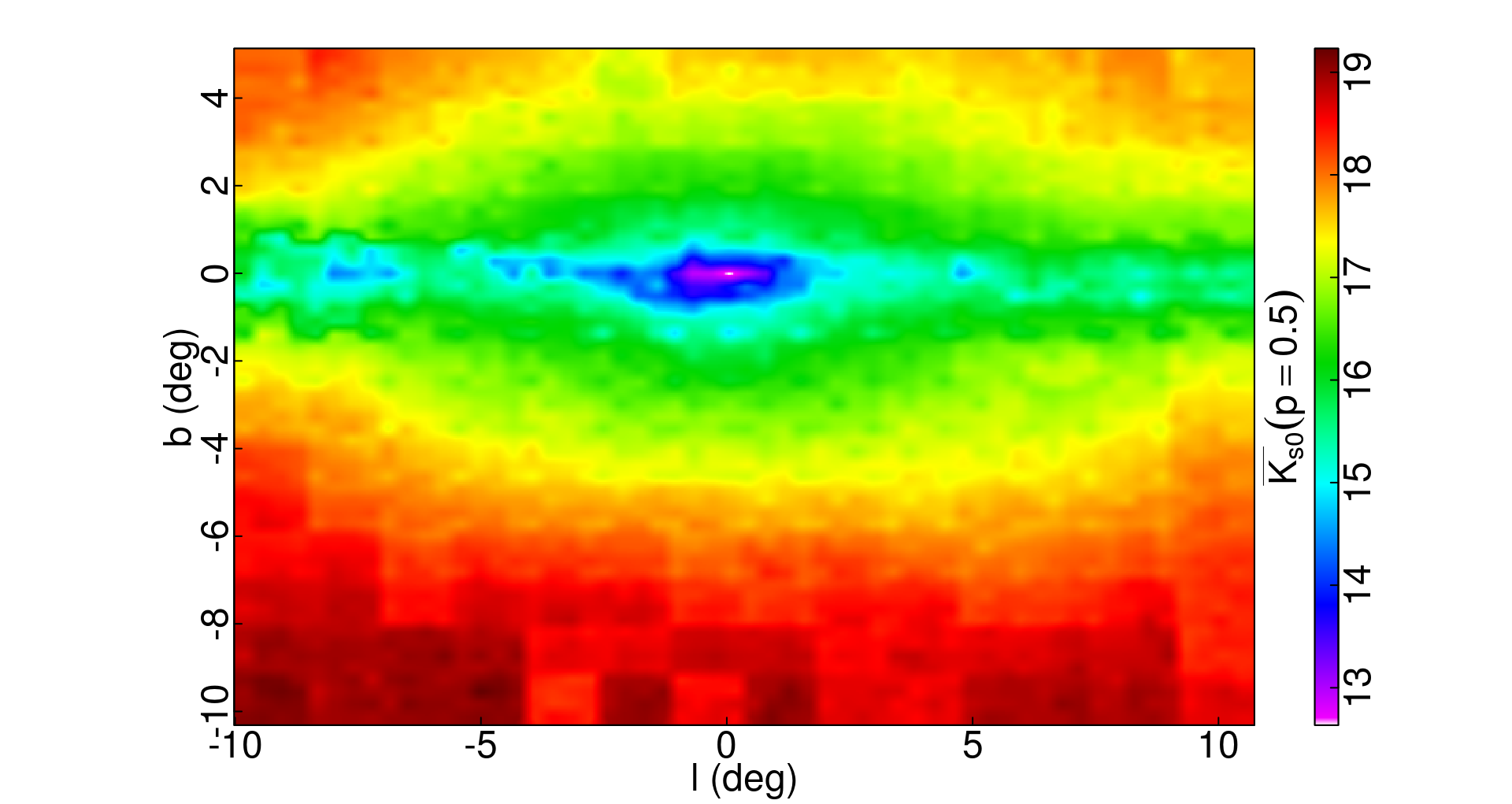}
	\caption[Color\--coded map of the mean $\Ks$ and $\mathrm{K_{s0}}$ of the stars in $p = 0.5$ levels across the whole bulge area in the dataset.]{Observed mean $\Ks$ (left panel) and reddening\--corrected $\mathrm{K_{s0}}$ magnitude of stars with $\sim$\,50\% completeness level $(p = 0.50 \pm 0.03)$ across the whole bulge area studied in this work. Each measurement comes from the average magnitude per individual detector per tile, where the position in the sky is set to be the middle of each detector field of view.}
	\label{fig:completeKs}
\end{figure*}

Figure~\ref{fig:completeKs} gives an overview of the global photometric completeness as a function of the position within the bulge, in the form of the mean $\Ks$ and $\mathrm{K_{s0}}$ \citep[$\Ks$ corrected for extinction from][]{surot+19} magnitudes of the stars at $50 \pm 3\%$ completeness level. 
By providing a global view of the photometric quality as a function of the star magnitude and position, the maps in Fig.~\ref{fig:completeKs} help to quickly understand the potential and usefulness of the photometry according to the type of studies that one is interested in.

For instance, for studies addressing the age determination a good sampling of the old MS\--TO region is needed.
Considering that for a 10\,Gyr old population of solar metallicity at the distance of the bulge (i.e. 8\,kpc), the MS\--TO is expected at $\Ks_0 \sim 17$, by looking at the reddening\--corrected completeness map (Fig.~\ref{fig:completeKs}, right panel), we can quickly assert that, with some exceptions at $l = \pm10^\circ$, any field within $|b| \lesssim 3.5^\circ$ is likely not complete enough for stellar dating.
Therefore, we should either discard or treat with particular caution any results derived from such regions. 
On the other hand, if the science goal is to study and make a census of the RC or bright RGB populations ($\Ks < 14$), then the photometric catalogs of all but the most central two tiles of the bulge are adequate. 

\begin{figure}  
	\centering
	\includegraphics[width=9 cm]{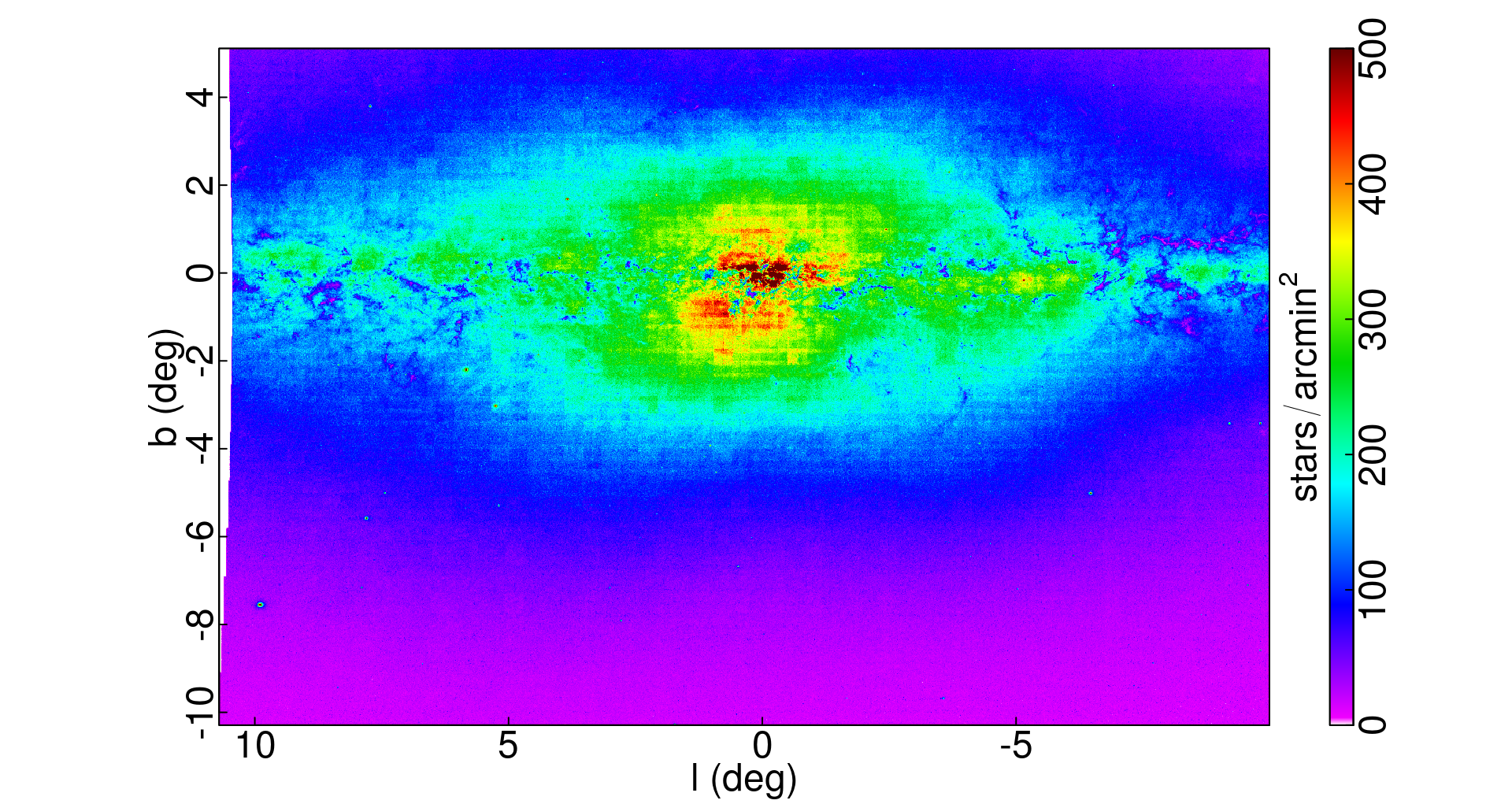}
	\caption{Stellar density map of the whole bulge area as derived by counting all stars with $\Ks < 16$. Duplicates have been addressed by counting stars in the first tile, while subsequent tiles are added excluding overlaps with adjacent fields.}
	\label{fig:starden}
\end{figure}

As expected the overall completeness of the derived photometry increases when moving from the center outwards because of the decreasing extinction and stellar density (i.e. confusion). 
In fact, the morphology of the bulge emerging from the completeness map closely resembles the boxy\--peanut shape that is pictured in the stellar maps shown in Fig.~\ref{fig:starden}.
Some Galactic globular clusters located in the VVV area reveal themselves in Fig.~\ref{fig:starden} as blue and cyan small dots because of their locally higher stellar density.

\begin{figure*}
	\centering
	\includegraphics[height=8cm]{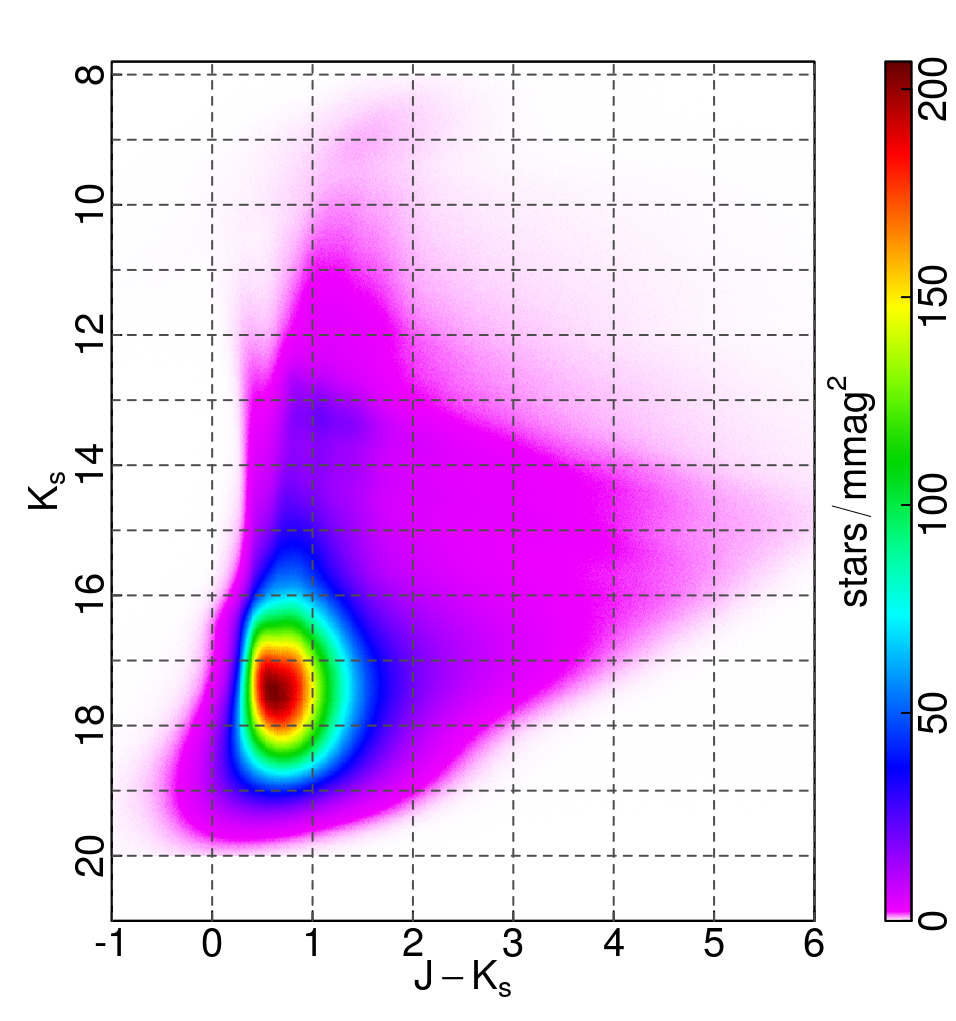}
	\includegraphics[height=8cm]{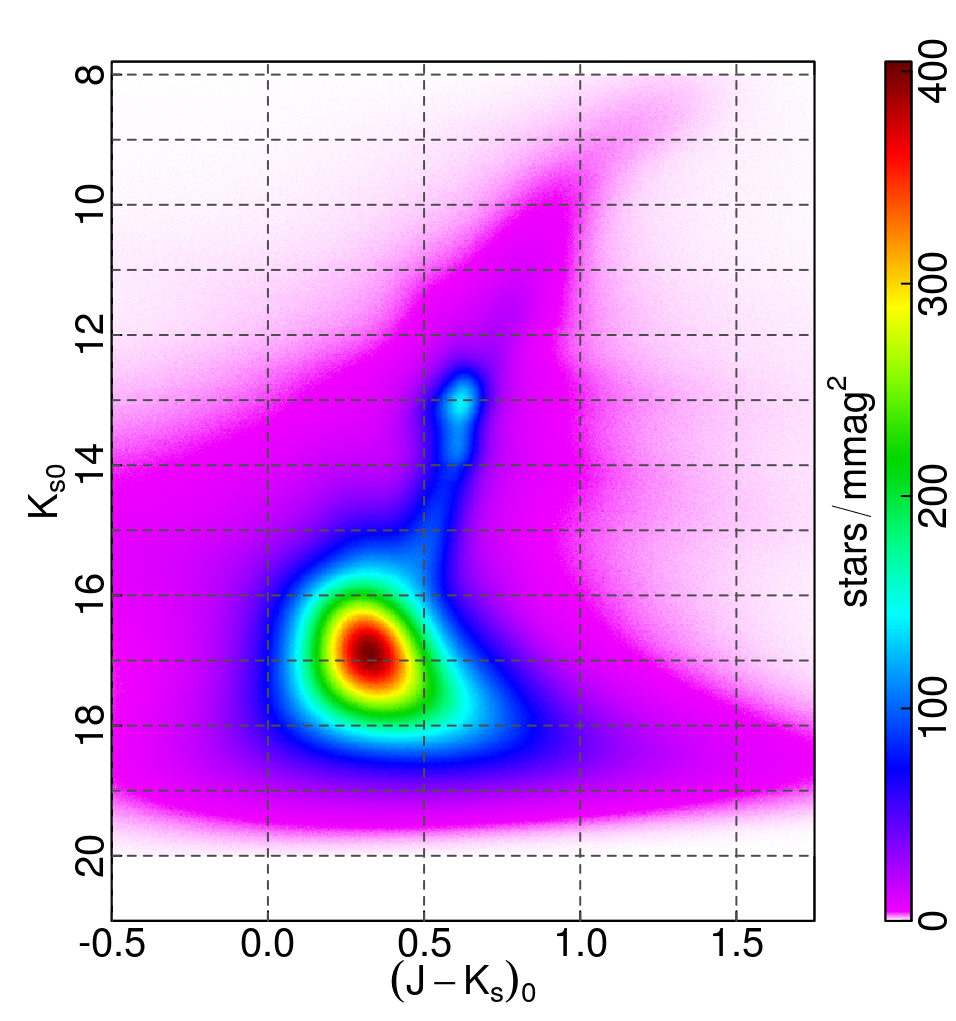}
	\caption[Hess diagram of the all 196 tiles.]{Hess diagram of all 196 tiles stacked together in the observed [$\Ks, \J-\Ks$] plane (left panel), and de\--reddened [$\Ks_0, (\J-\Ks)_0$] plane (right panel) by using the color\--excess map and the extinction law from \cite{surot+19}. The CMDs comprise nearly 600 million stars.}
	\label{fig:supercmd}
\end{figure*}

To provide a global and quick view of the stellar content within the VVV area, in Fig.~\ref{fig:supercmd} we show the CMD obtained by stacking together all 196 tiles (i.e. from b201 to b396) with $\sim$\,580 million sources, but not accounting for the $\sim$\,10\% global overlap between tiles.
In the [$\Ks\,vs\,(\J-\Ks)$] plane (left panel of Fig.~\ref{fig:supercmd}) the effect of the large reddening is clearly evident, smearing the stars' color over a huge range (i.e. $-1 \lesssim (\J-\Ks) \lesssim 6$), and making the RGB and RC sequences very blurred and hard to identify.
On the other hand, when we correct the photometry by using the reddening map and the extinction law from \cite{surot+19} shown in right panel of Fig.~\ref{fig:supercmd}, the sequences of the evolved bulge stellar populations stand out clearly.
We note that the color spread in the blue part of the CMD at $\Ks_0\lesssim16$ is not representative of the actual spread in colors of disk stars because it is the result of applying the bulge extinction correction to the bright MS disk stars along the line\--of\--sight that, unlike background bulge (and a portion of the disk) stars, are not affected by the entire amount of intervening dust. In other words, the large extinction is confined mostly within the bulge only and not between us and the disk as shown in 3D extinction maps \citep[e.g.,][]{schultheis14,chen19}.

Finally, we reckon that because of the large differential extinction and distance depth, for detailed study of the stellar populations along the bulge line\--of\--sight one should not use Fig.~\ref{fig:supercmd}, but instead consider smaller CMDs obtained within smaller sub\--regions.

\section{Applications}
To demonstrate the tremendous potential inherent to this new photometric dataset in what follows we present examples of possible applications.

\subsection{I. Tracing the RC stars distribution}
\label{sec:app-rc}
As well know since long time, RC stars are very useful standard candles thank to the fact that their magnitude variation as a function of the age and metallicity is accurately predicted by stellar evolution models \citep{salaris+02}.
In addition, because RC are rather bright stars, and as such they are easily recognizable in the CMD plane (especially in the near\--IR domain), over the years they have been extensively used for a variety of studies aimed at: {\it i)} characterizing the bulge morphology and structure \citep[see][and references therein]{wegg+13,valenti+16}, {\it ii)} deriving reddening maps \citep[i.e.][and reference therein]{oscar12}, and {\it iii)} addressing the bulge age \citep[][i.e. the RC distribution provides an anchor when comparing CMDs of different fields]{zoccali+03, valenti+13, surot+18}.
\begin{figure*}
	\centering
	\includegraphics[height=9cm]{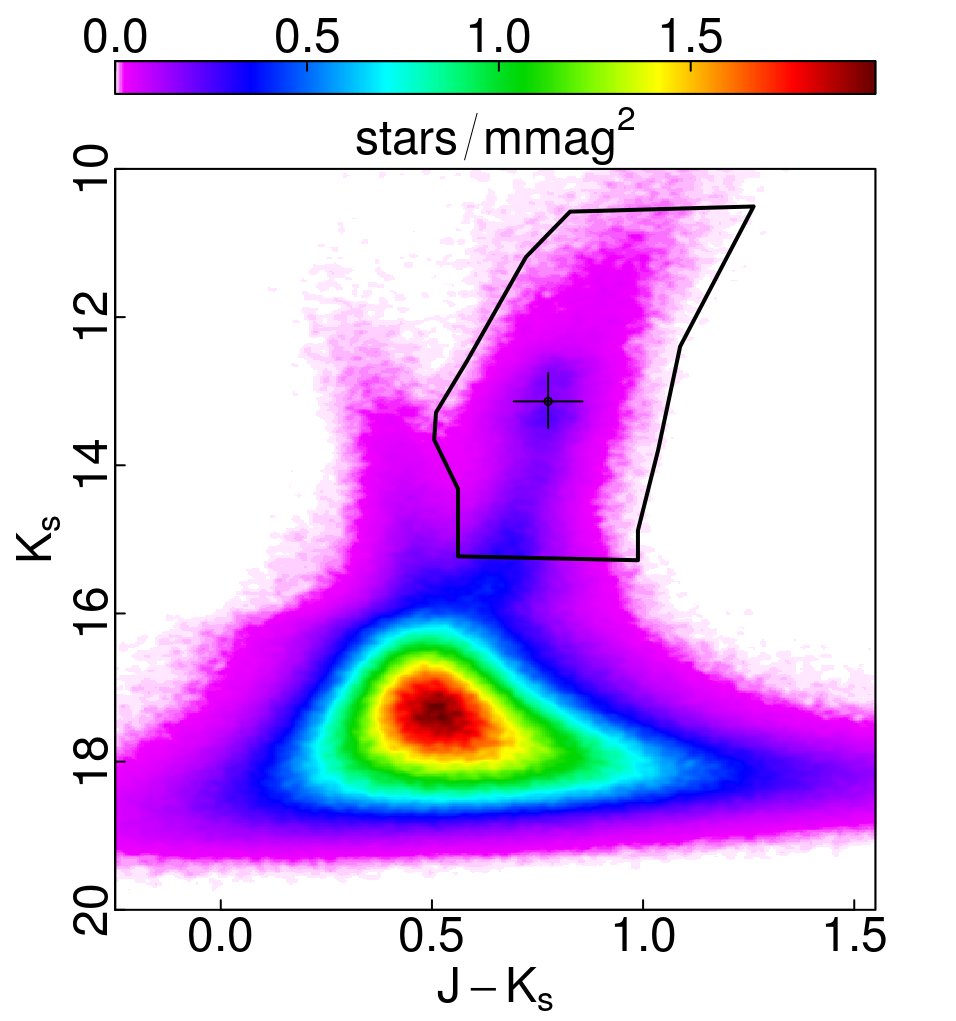}
	\includegraphics[height=8cm]{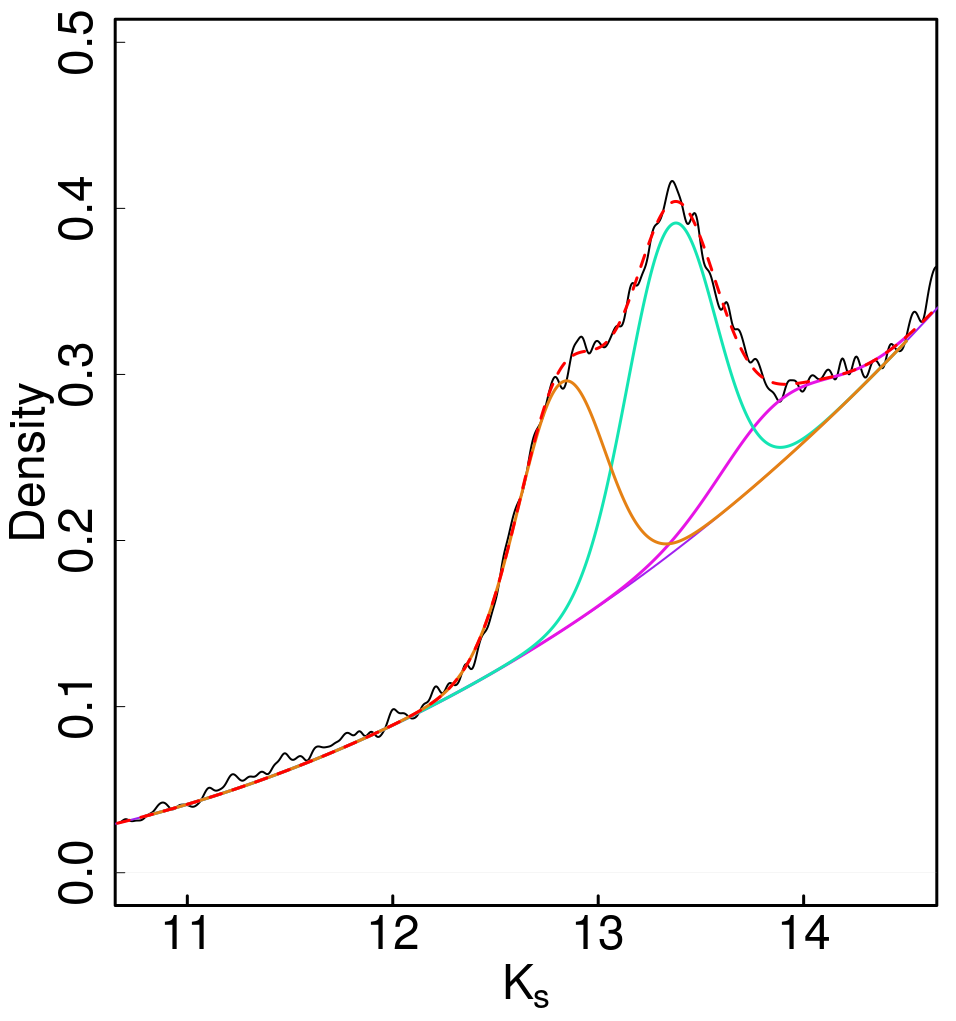}
	\caption{Example of the adopted procedure to trace the RC distribution in a typical field where the X\--shape of the bulge is detected (see text). {\it Left panel:} observed Hess diagram of b249 tile, with a black polygon acting as the selection window used to isolate RC and RGB stars, and to construct the corresponding luminosity function reported in the right panel. The black cross shows the final estimate of the mean RC position, together with its effective width in color and magnitude. {\it Right panel:} kernel estimate of the selected RC and RGB luminosity function, not corrected for completeness (black solid line). Also plotted are the fit to the RGB distribution (violet line), and three gaussian components used to fit the RCs (apricot and cyan) and RGB\--bump (magenta). The global fit using all components is shown as  dashed red line. In this example, the mean and $\sigma$ of the gaussians component are: $\Ks=12.82 \pm 0.21$\,mag, $13.36 \pm 0.22$\,mag, for the RC, and $13.86 \pm 0.28$\,mag for the RGB\--bump. In this exercise, no reddening nor completeness correction has been applied.}
	\label{fig:b249_rc}
\end{figure*} 

This accurate PSF photometry enables tracing the RC population across the whole VVV area. 
To detect the mean magnitude of the RC population across the surveyed area we have adopted a procedure very similar to what previously done by \citet{nataf+11,oscar12, wegg+13,valenti+16}.
Specifically, we start by making an educated color\--magnitude cut in the CMD such as to include evolved stars only (e.g. RGB, and RC) down to the magnitude and color levels where their sequence is still distinguishable from the bright MS of the disk (see left panel in Figure~\ref{fig:b249_rc}). 
Then we estimate the $\Ks$\--magnitude star density with a narrow enough gaussian kernel as to not introduce extra dispersion (i.e. very similar to luminosity function histogram, but without depending on the arbitrary choice of a starting bin position). 
We fit the so\--derived luminosity function with a $4^{th}$\--degree polynomial\footnote{$2.14-0.635x+0.0701x^2-0.00362x^3+8.34*10^{-5} x^4$} by excluding the magnitude region around the RC, where there is evident contribution from any localized overdensities (i.e. RCs and RGB\--bumps).
By subtracting the fit to the RC region, the residual corresponds to the distribution of RC and RGB\--bump stars only.
Finally, to trace the RC population in term of mean magnitude and star counts we fit the residual with gaussians (see right panel in Figure~\ref{fig:b249_rc}).

Because of the bulge X\--shape, we know that some fields ($|b|>4^\circ, |l|\lesssim4^\circ$) show a prominent bimodality in the RC profile, therefore in those regions we have opted to use at least 3 gaussians to fit the RC profile: two for the RC and one for the RGB\--bump.
It is worth mentioning that in principle, in the fields showing the split of the RC we should expect also 2 RGB\--bumps. 
However, because the luminosity of RGB\--bump corresponding to the brightest RC (i.e. the southern arm of the X\--shape closer to us) overlaps with the luminosity range spanned by the second and fainter RC, its detection and proper characterization is usually very hard. In fact, as well predicted by stellar evolution theory \citep{renzini+88} the RGB\--bump is considerably less populated than the RC. \citet{nataf+11} estimates that in the bulge the RGB bump population makes $\lesssim$\,30\% of the RC stars.

\begin{figure}
\centering
\includegraphics[height=5cm, trim = {7cm 0.2cm 7cm 1.9cm}, clip]{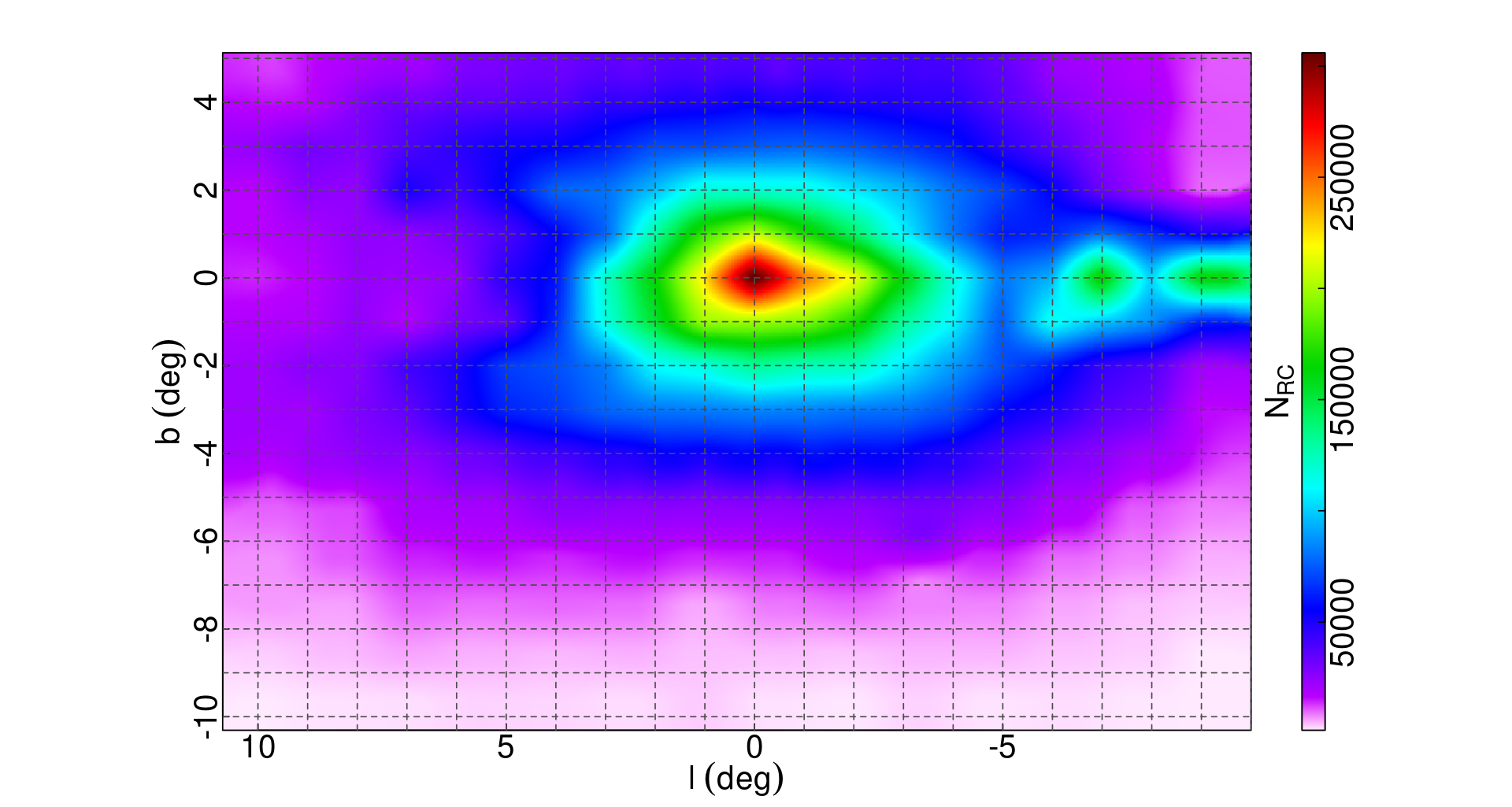}
\caption{Density map in the Galactic longitude\--latitude plane based on RC star counts. Star counts for the region at $|b|\leq4^\circ$ are taken from \cite{valenti+16}, whereas from this study for the outer region.}
\label{fig:rcmap}
\end{figure}

\begin{figure}
\centering
\includegraphics[height=7cm]{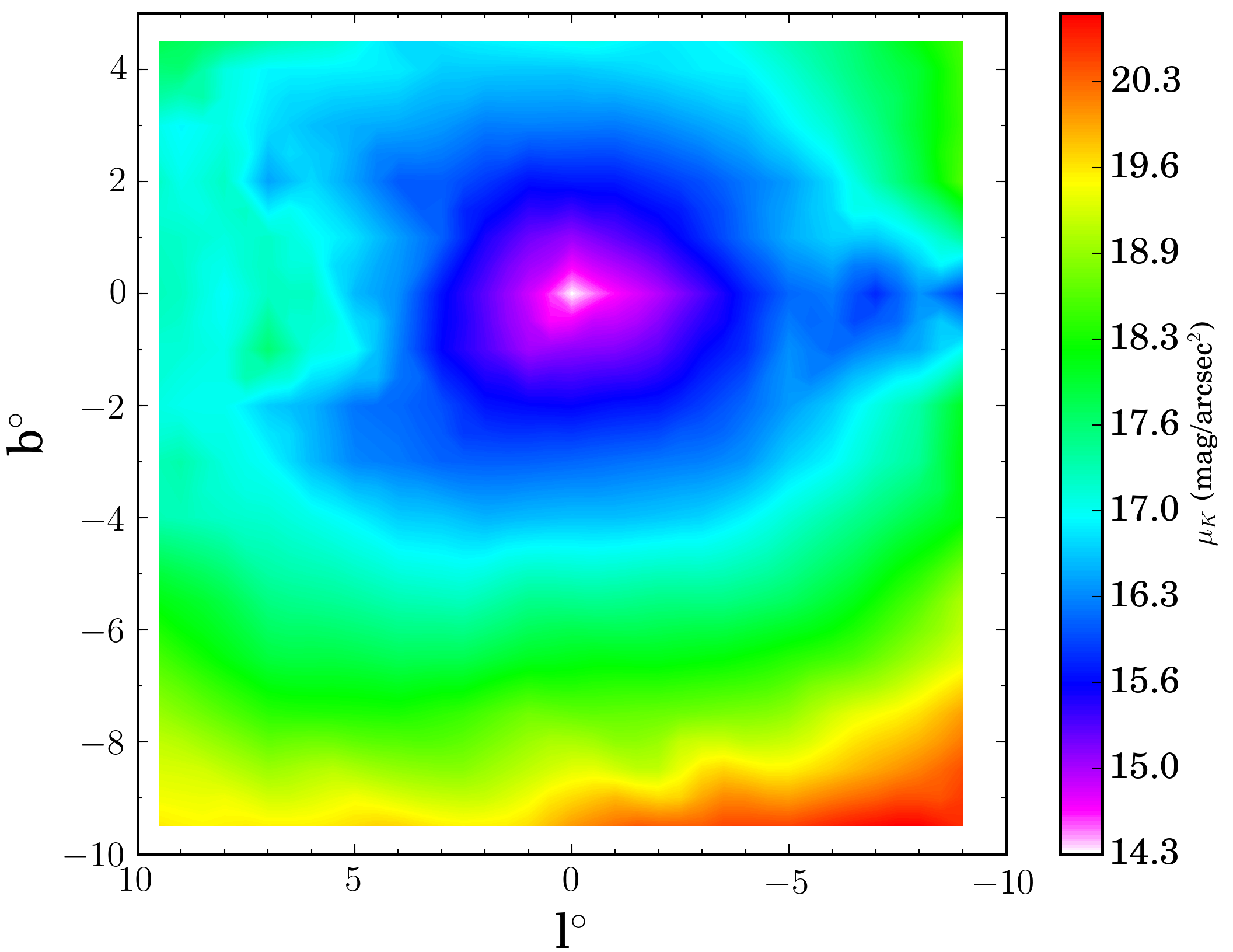}
\caption{Intrinsic (i.e. reddening corrected) surface $\Ks$ magnitude map in the Galactic longitude\--latitude plane.}
\label{fig:sbmap}
\end{figure}

To provide a couple of examples of possible uses of the RCs, enabling the study of the global bulge structure and morphology, here we show the stellar density (see Figure~\ref{fig:rcmap}) and surface magnitude (see  Figure~\ref{fig:sbmap}) maps.
The stellar density map is derived by combining the RC distribution from \cite{valenti+16} (for the region $|b|\leq4^\circ$) and from this study (for the outer regions). 
As described in \citet{valenti+16}, the bulge density map clearly shows the expected boxy/peanut morphology, with increasing stars density towards the centre. 
In addition, the observed star counts asymmetry along the bulge minor axis (i.e. higher density at negative longitudes) is consistent with the presence of a bar that has its close side at positive longitude.

The surface magnitude profile in Figure~\ref{fig:sbmap} is obtained by using the RC star counts shown in Fig.~\ref{fig:rcmap} and  the mean intrinsic magnitude of the RC at any given line\--of\--sight (i.e. $\Ks_0^{RC}$ observed in each tile), under the assumption that for complex stellar systems older than 2\--3 Gyr, the contribution of the RC population to the total luminosity is 12\% \citep{maraston+05}.

Here the RC stars are used as tracers of the intermediate\--to\--old population, that should exclude most of the disk contamination, effectively producing maps that trace instead the density of stars and surface magnitude profile in the bulge alone. 

\subsection{II. Gaia vs VVV}
\label{sec:app-gaia}
The cross\--match between this VVV photometry and Gaia DR2 \citep{gaiaDR2} catalogs potentially allows to combine the photometric and kinematic (i.e. proper motions) properties of the stellar population over most of the bulge extension.
However, in practice because of the extinction and the crowding in some bulge regions Gaia data is severely limited.

\begin{figure*}
\centering
\includegraphics[width=16cm]{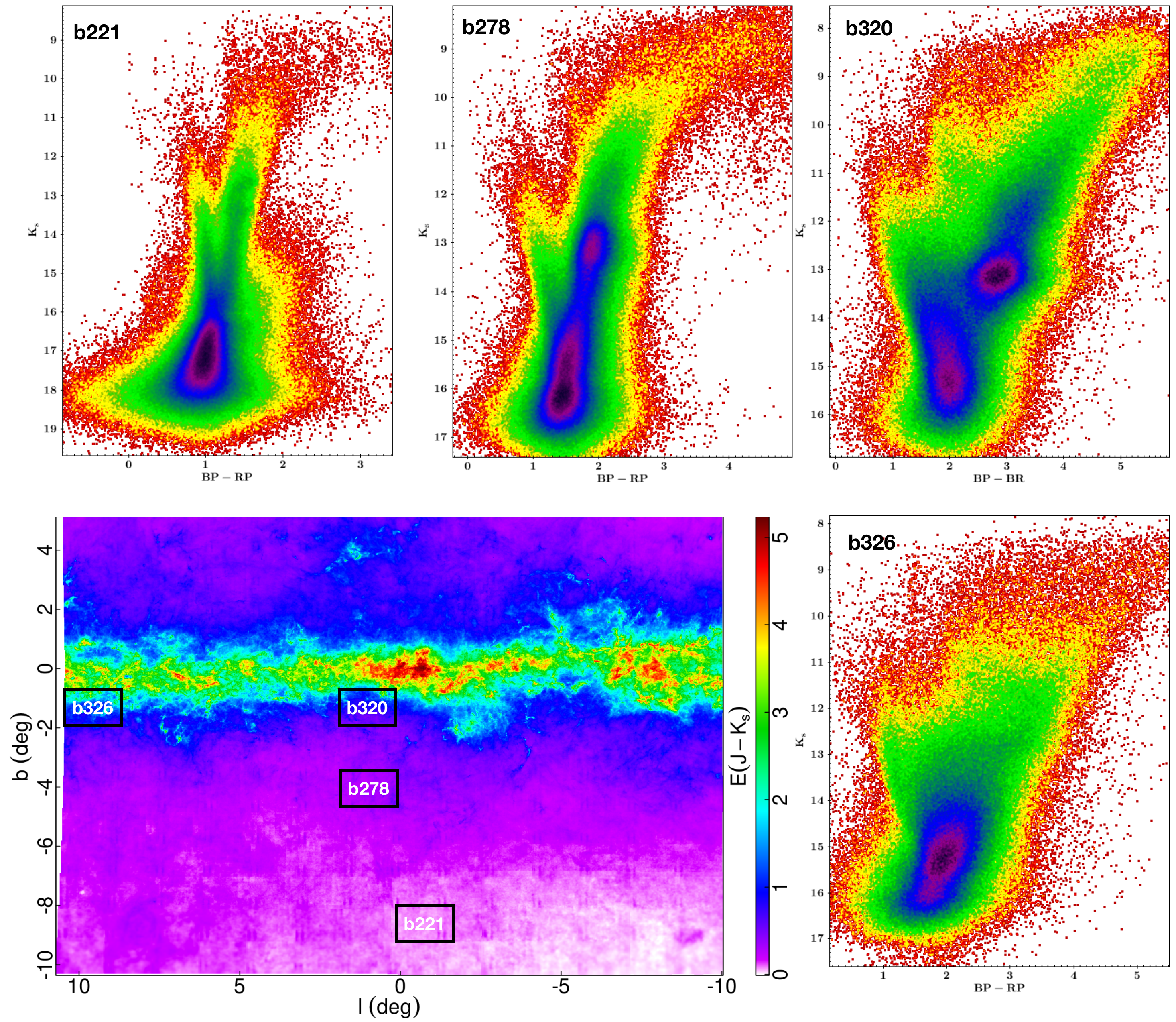}
\caption{CMDs of 4 VVV fields in the [$\Ks, BP-BR$] plane as obtained by cross\--matching the Gaia DR2 and this new VVV photometry (upper and lower right panel). The position of the 4 fields is displayed on the new bulge reddening map of \citet[][lower left panel]{surot+19}. }
\label{fig:gaia}
\end{figure*}

In what follows, by using the photometric completeness of our catalogs, we assess how well Gaia is able to sample the bulge stellar population.
Specifically, in Fig.\,\ref{fig:gaia} we show the combined Gaia\--VVV CMDs of 4 selected fields representative of different  level of extinction and crowding.
In the outer bulge regions (see CMD of tile b221 in Fig.\,\ref{fig:gaia}), Gaia data samples the bulge population down to $\sim$\,1.5\,mag below the old MS\--TO (i.e. $\Ks\sim17$), hence one can use the stars proper motions to  separate the bulge component from the intervening disk population along the line of sight allowing to date the bulge stars by means of the MS\--TO luminosity.
At the latitude of Baade's Window (i.e. $b\sim-4^\circ$, see CMD of tile b278 in Fig.\,\ref{fig:gaia}), the Gaia photometry is already no longer deep enough to enable any kind of age study being highly incomplete at the MS\--TO level.
On the other hand, the RC bulge stellar population is properly sampled by Gaia throughout most of the VVV area with the exception of the innermost $\sim\pm1.5^\circ~$ across the Galactic plane.
At $b\sim-1^\circ$, the quality of the Gaia photometry is highly variable because of both crowding and reddening.
Only when the latter is still relatively small (i.e. for $E(\J-\Ks)\lesssim1.5$), Gaia is able to samplee the RC population.
This is clearly evident when one compares the CMD of the two fields b320 and b326 (see Fig.\,\ref{fig:gaia}), which turn out to be very different despite the fields being located at the same latitude.
Although b326 is in a much less crowded region than b320 (see the RC star density map in Fig.\,\ref{fig:rcmap}), the combined Gaia\--VVV CMD of b326 barely samples the RC distribution because of the large reddening. Noteworthy is, that the limitations of Gaia in this regions could be improved dramatically with LSST mapping the bulge \citep{oscarlsst}.

\section{Summary}
We provide a detailed and comprehensive census of the Milky Way bulge's stellar populations by  releasing new accurate photometric catalogs of the central $\sim$\,300\,deg$^2$.
The photometric database, containing nearly 600 millions stars, consists of 196 catalogs obtained by performing PSF\--fitting photometry of multi\--epoch J and K$_s$ VVV images.
Extensive artificial star experiments conducted on all 3912 images allowed to properly assess the completeness and accuracy of the photometric measurements.
With limiting magnitudes $\Ks,\approx20$ and $\J\approx21$, this new photometric compilation allows to characterize the evolved and un\--evolved stellar population of the bulge over most of its extension.
In particular, the RC stellar population is properly sampled with a photometric completeness ranging from nearly 100\% to 70\% throughout the VVV bulge area, with the exception of the innermost field close to the Galactic center where the completeness drops to 50\%.

The photometry is accurate and deep enough to sample the old MS\--TO across the whole outer bulge region (i.e. $|b|\gtrsim3.5^\circ$) with over $\sim$\,50\% completeness, hence enabling studies of stellar ages, and star formation history reconstruction based on synthetic CMD\--fitting techniques \citep[see e.g.][]{surot+18}.

Finally, the entire photometric dataset is publicly available to the whole community at the ESO Science Archive http://archive.eso.org/scienceportal/home.

\begin{acknowledgements}
We thank the Referee for his/her comments, which helped us to improve the quality of the paper.
FS, EV and MR acknowledge the support by the Excellence Cluster "Origins and Structure of the Universe". 
The simulations have been carried out on the computing facilities of the Computational Center for Particle and Astrophysics (C2PAP). 
Support for MZ and  DM  is  provided by  the  BASAL  CATA  Center  for Astrophysics  and  
Associated Technologies through grant  AFB-170002, and the Ministry for the Economy, 
Development, and Tourism's Programa Iniciativa  Cient\'\i fica Milenio through grant IC120009, 
awarded to Millenium Institute of Astrophysics (MAS). 
MZ acknowledge support from FONDECYT Regular 1150345.
DM acknowledges support from FONDECYT Regular 1170121.
This research made use of Astropy, a community-developed core Python package for Astronomy \citep{astropyI,astropyII}., R: A language and environment for statistical computing \citep{R}, as well as several packages (akima: \citealt{akimaR}, gplots: \citealt{gplotsR}, spatstat: \citealt{spatstat})

\end{acknowledgements}
\bibliographystyle{aa}
\bibliography{myrefs}

 \appendix
 \section{VVV stacked pawprint images}
\begin{table*}
	\caption{List of the VVV stacked pawprint images used in this work, together with the corresponding image quality, airmass and ellipticity as derived by the fits header. The full table is electronically available at CDS, and at the ESO \textbf{Science Archive}}    
	\label{tab:allima}      
	\centering          
	\begin{tabular}{ l c c c c c }     
		\hline    
		Image name$^a$ & Field$^b$ & Filter & IQ & airmass & Ellipticity\\ 
		  &  &   & arcsec &  \\ 
		  \hline
v20100411\_01136\_st & b201 & J & 1.053 &0.81 & 0.10\\
v20100411\_01138\_st & b201 & J & 1.052 & 0.80 & 0.11  \\
v20100411\_01140\_st & b201 & J & 1.052 & 0.82 & 0.11  \\
v20100411\_01142\_st & b201 & J & 1.051 & 0.78 & 0.13  \\
v20100411\_01144\_st & b201 & J & 1.051 & 0.77 & 0.13  \\
v20100411\_01146\_st & b201 & J & 1.051 & 0.81 & 0.09  \\
v20150822\_00534\_st & b201 & J & 1.204 & 0.75 & 0.06  \\
v20150822\_00536\_st & b201 & J & 1.207 & 0.71 & 0.06  \\
v20150822\_00538\_st & b201 & J & 1.210 & 0.69 & 0.06  \\
v20150822\_00540\_st & b201 & J & 1.210 & 0.73 & 0.05  \\
v20150822\_00542\_st & b201 & J & 1.212 & 0.73 & 0.06  \\
v20150822\_00544\_st & b201 & J & 1.213 & 0.68 & 0.07  \\
v20110516\_00680\_st & b201 & K$_s$ & 1.048 & 0.51 & 0.07  \\
v20110516\_00682\_st & b201 & K$_s$ & 1.048 & 0.52 & 0.06  \\
v20110516\_00684\_st & b201 & K$_s$ & 1.048 & 0.54 & 0.07  \\
v20110516\_00686\_st & b201 & K$_s$ & 1.047 & 0.52 & 0.06  \\
v20110516\_00688\_st & b201 & K$_s$ & 1.047 & 0.52 & 0.07  \\
v20110516\_00690\_st & b201 & K$_s$ & 1.047 & 0.51 & 0.07  \\
v20110513\_00711\_st & b201 & K$_s$ & 1.134 & 0.53 & 0.08  \\
v20110513\_00713\_st & b201 & K$_s$ & 1.136 & 0.53 & 0.06  \\
v20110513\_00715\_st & b201 & K$_s$ & 1.137 & 0.52 & 0.06  \\
v20110513\_00717\_st & b201 & K$_s$ & 1.137 & 0.53 & 0.06  \\
v20110513\_00719\_st & b201 & K$_s$ & 1.137 & 0.55 & 0.06  \\
v20110513\_00721\_st & b201 & K$_s$ & 1.138 & 0.57 & 0.07  \\
		\hline
		\hline \hline
	\multicolumn{4}{l}{$^{a}$ Name of the image file as given by the CASU pipeline}\\
	 \multicolumn{4}{l}{$^{b}$ VVV field name}\\
	\end{tabular}
\end{table*}

\end{document}